\documentclass[aps,preprint]{revtex4}
\usepackage{amsfonts}
\usepackage{amsmath}
\usepackage{amssymb}
\usepackage{graphicx}
\usepackage{subfigure}

\newtheorem{theorem}{Theorem}
\newtheorem{acknowledgement}[theorem]{Acknowledgement}

\begin{document}

\title[ ]{Applications and identification of surface correlations}
\author{M. Escobar, A. E. Meyerovich \thanks{%
Corresponding author: A. E. Meyerovich, meyerovich@phys.uri.edu, Phone:
+1.401.874.2047, FAX: +1.401.874.2380}}
\affiliation{Department of Physics, University of Rhode Island, Kingston, RI 02881-0817,
USA}
\author{}
\affiliation{}

\begin{abstract}
We compare theoretical, experimental, and computational approaches to random
rough surfaces. The aim is to produce rough surfaces with desirable
correlations and to analyze the correlation functions extracted from the
surface profiles. Physical applications include ultracold neutrons in a
rough waveguide, lateral electronic transport, and scattering of longwave
particles and waves. Results provide guidance on how to deal with
experimental and computational data on rough surfaces. A supplemental goal
is to optimize the neutron waveguide for GRANIT experiments. The measured
correlators are identified by fitting functions or by direct spectral
analysis. The results are used to compare the calculated observables with
theoretical values. Because of fluctuations, the fitting procedures lead to
inaccurate physical results even if the quality of the fit is very good
unless one guesses the right shape of the fitting function. Reliable
extraction of the correlation function from the measured surface profile
seems virtually impossible without independent information on the structure
of the correlation function. Direct spectral analysis of raw data rarely
works better than the use of a "wrong" fitting function. Analysis of
surfaces with a large correlation radius is hindered by the presence of
domains and interdomain correlations.
\end{abstract}

\volumeyear{year}
\volumenumber{number}
\issuenumber{number}
\eid{identifier}
\date[Date text]{date}
\received[Received text]{date}
\revised[Revised text]{date}
\accepted[Accepted text]{date}
\published[Published text]{date}
\startpage{1}
\endpage{}
\maketitle

\section{Introduction}

Progress in material science, nanofabrication and related technologies
expands the range of physical systems in which scattering by surface and
interface roughness is the dominant scattering channel. Such systems are
studied by different theoretical, experimental, and computational
techniques, all of which, in principle, should use a more or less common
language and converge to identical results. Below we try to answer the
question how to bridge the gap between these techniques. Our applied goal is
to find ways to prepare a random rough surface with desirable physical
properties.

We are interested in surfaces with slight random roughness for which the
observables are quadratic in roughness. Theoretical expressions for the
physical observables, such as, for example, transport coefficients, should
explicitly contain the geometrical and statistical parameters of surface
roughness. These parameters are routinely introduced (see, \emph{e.g.,} Ref. 
\cite{q2} and references therein) by the binary roughness (auto-)correlation
function $\zeta \left( x\right) ,$ which is usually characterized by an
average amplitude of inhomogeneities $\ell $ and a single correlation radius
of inhomogeneities $R$, 
\begin{equation}
\zeta \left( x\right) =\ell ^{2}\varphi \left( x/R\right) ,\ \varphi \left(
0\right) =1.  \label{i1}
\end{equation}%
An equivalent description uses the roughness structure function, $S\left(
x\right) =\ell ^{2}\left( 1-\varphi \left( x\right) \right) $. A brief
review of alternative approaches to roughness can be found in Ref. \cite%
{bernasek1}. For applications, the Fourier image of the correlation function 
$\left( \ref{i1}\right) $ (the so-called power spectrum of surface
roughness),%
\begin{equation}
\zeta \left( q\right) =\ell ^{2}\psi \left( qR\right) ,  \label{ii1}
\end{equation}%
is often more important than the correlation function itself (here $q$ is an
appropriate conjugate for $x;$ in \textrm{1D} there could be an extra
coefficient $\sqrt{2\pi }R$, in \textrm{2D} - just $R^{2}$). The use of
multiparameter descriptors instead of Eqs. $\left( \ref{i1}\right) ,\left( %
\ref{ii1}\right) $ could provide additional fitting parameters, but usually
does not clarify the physics.

The form of the roughness correlation function for real surfaces cannot be
predicted theoretically except for a few exactly solvable models of surface
interaction which may or may not correspond to reality. Even the simplest
models rarely lead to simple explicit expressions for the correlation
functions. One can also try to establish classes of universality for
roughening and to find the roughening or fractal exponents (for recent
examples see Refs. \cite{fracture1,fracture2,fracture3} and references
therein). In theoretical calculations the correlation function is usually
assumed to be known leaving its determination to experiment or numerical
modeling.

These sources are often inconclusive and the theoretical evaluation of
observables is performed using some \textrm{ad hoc} correlation function.
The variety of Gaussian, exponential, or power law correlators are used
almost at will despite the evidence that the choice of the correlation
functions with similar correlation parameters but of different functional
forms can lead to very different physical results (see, \emph{e.g.,} Refs. 
\cite{q2,fer1,munoz1,pon1}). All this degrades the application of
theoretical results to real surfaces.

Thus the questions are whether it is possible to extract an accurate
correlation function from experiment and whether it is possible to create a
random rough surface with a predetermined roughness correlation function. We
will start from the former question and later give a physical example for
which the latter question is indeed crucial.

There are two types of experiments which can provide information on the
surface correlation function: scattering of particles or waves by the rough
surface and direct measurements of the surface profile. The intensity of
waves scattered from a rough surface is directly described by the power
spectrum of the correlation function $\left( \ref{ii1}\right) $ \cite{q2},
but the accuracy of measurements is high only in limited ranges of wave
vectors and angles.

The second type of experiment seems more promising since one can easily
extract the correlation function $\zeta ^{\exp }$ from precise scanning
measurements of the surface profiles, such as, for example, STM or AFM. The
difficulty here lies in proper identification of the raw data on $\zeta
^{\exp }$. This extracted discrete correlation function $\zeta ^{\exp }$
inevitably exhibits noticeable noise, especially if the scanned area is not
very large, and cannot be unambiguously identified with some simple
functional form of $\zeta \left( x\right) $, Eq. $\left( \ref{i1}\right) $.
There are two ways of dealing with these difficulties: either compare the
extracted correlator $\zeta ^{\exp }$ with some preconceived fitting
function $\zeta ^{fit}\left( x\right) $ and get the correlation parameters
from the best fit or feed the the extracted raw correlation function $\zeta
^{\exp }$ directly into the theoretical equations for observables. Below we
analyze the limitations of both approaches.

The experimental difficulties of extracting an accurate surface correlator
multiply when one deals with an atomic-scale roughness, even if one
disregards the issue of the accuracy of the data on the surface profile
related, for example, to the tip profile \cite{stm1} or the step size \cite%
{munoz4}. Some requirements on accuracy of profile measurements for reliable
extraction of the correlation parameters are discussed in Refs. \cite%
{ogilvy2,shikin1}. It is not even clear to what extent the theoretical
methods using the correlation function of surface roughness can be applied
to random inhomogeneities on atomically-smooth crystal surfaces.

The potential shortcomings of the first approach are obvious: the
correlation parameters which are extracted from $\zeta ^{\exp }$ in this
way, depend on an \textrm{ad hoc }choice of the fitting function. Fitting of
the STM data on $\zeta ^{\exp }\left( x\right) $ to correlation functions $%
\zeta ^{fit}\left( x\right) $ of different functional forms could yield
vastly different values of the correlation parameters such as the
correlation radii $R$ (for recent experimental examples see, e.g., Refs. 
\cite{munoz1,munoz2}). This can become a real problem when the step size in
scanning microscopy is comparable to the correlation radius of roughness:
according to the estimates \cite{q2}, to resolve the shape of the
correlation peak one needs about ten points within the peak. As we will see,
even the increase in the sample size does not necessarily help. In the end,
using a preconceived correlation function is especially dangerous in two
limiting cases when the correlation radius $R$ is comparable to the scanning
step or when $R$ and, therefore, the size of inhomogeneities, is large.
However, as we will see below, the use of the raw experimental data on $%
\zeta ^{\exp }$ can often be more dangerous than the risk of using the wrong
fitting function.

But how can one evaluate the reliability of identifying the correlation
function extracted from precise scanning measurements of the surface
profile? As we will see, the statistical quality of a fit to some fitting
function is not the answer.

The main issue is that we cannot fabricate a surface with a known
correlation function to serve as a reference to check against the extracted
correlator. What we can do instead is to computationally generate a surface
with a given correlation function, scan this surface, and analyze the
extracted correlators. The knowledge of the exact correlation function will
allow us to judge the quality of identification not by statistical
properties of the fits, but by how well the physical observables are
reproduced. The identification issues for real and computationally generated
surfaces are more or less the same \cite{goodnick1,shikin1,palas1,munoz2}
and our results should provide a roadmap for dealing with experimental data.
This will also allow us to accomplish our second applied goal: to design a
random surface with desirable correlation properties which in the case of a
reasonable physical scale can be reproduced experimentally (see below).

We start (Section II) from two computational procedures for generating
random rough surfaces with known correlation functions. The first procedure
produces a random rough surface with any predetermined correlation function
and is suitable for larger scale roughness. The second one relies on model
Hamiltonians. It provides surfaces with discretized profiles, more
appropriate for atomic-scale roughness, but with a limited number of
correlators. Which of the procedures is preferable depends on the physical
circumstances.

In Section III we briefly describe physical applications which we use to
test the results. For clarity, we chose the applications for which the
roughness contribution to the observables collapses into a single constant.
This ensures effective and unambiguous evaluation of the quality of the data
and our methods. The first of these applications, namely, quantized
ultracold neutrons in rough waveguides, is essentially a one dimensional (%
\textrm{1D}) application with a large spatial scale of roughness (the
typical scale is 6 $\mathrm{\mu m}$). Here we also have a practical goal:
designing the best rough waveguide for experiments in GRANIT installation
(ILL, Grenoble) by optimizing the waveguide roughness. Transferring the
generated profile onto the real mirror seems to be technically feasible
because of a large spatial scale of roughness and is by far preferable to
the current procedure of introducing the uncontrolled roughness (random
scratching of the mirror). Our second application is more traditional and
deals with the conductivity of two dimensional (\textrm{2D}) ultrathin films
in quantum size effect (QSE) conditions and, more generally, with scattering
of longwave particles and waves by rough surfaces.

In Section IV we analyze random surface profiles generated using the methods
of Section II. We extract the correlation functions from these profiles and
try to identify them by fitting to different types of the fitting functions
using the same procedures used in analyzing the results of the scanning
microscopy measurements with a fixed step. The results of the fits are then
used to calculate the observables for the applications from Section III. The
purpose here is to find out what kind or errors are introduced by \textrm{ad
hoc} assumptions about the shapes of the fitting functions when analyzing
experimental and numerical data on surface profiles. Since in this case we
know the "true" correlation functions, we have an excellent criterion to
compare the errors. We will see that the quality of the fit, which is
described by the standard deviation $\sigma $ between $\zeta ^{\exp }\left(
x\right) $ and the fitting function $\zeta ^{fit}\left( x\right) $ does not
translate into the quality of the physical results unless the fitting
function $\ell ^{2}\varphi ^{fit}\left( x/R\right) $ has the right
functional form which is, unfortunately, unknown in most experiments with
real surfaces. In many cases the physical results turn out even worse if one
tries to input the raw experimental data on $\zeta ^{\exp }$\ directly into
the calculations instead of risking to make a wrong guess about the
functional form of the fitting function $\varphi ^{fit}\left( x/R\right) $.
The results are summarized in Section V.

\section{Generation of rough surfaces}

In this Section we briefly describe two numerical methods for generating
random rough surfaces with predetermined correlation functions. A short
review of alternative approaches is given, for example, in Ref. \cite%
{tribology1}. Some of the earlier work in this direction can be found in
Refs. \cite{ogilvy2,navarini1}.

\subsection{Surfaces with arbitrary correlation functions}

In this subsection we generate random surfaces with an arbitrary
predetermined correlation function of surface roughness without paying
attention to discretization of the amplitudes on atomic scale. In this
sense, we will be generating macroscopic or "classical" roughness with the
only constraint that the profiles are described by the smooth functions.
This is appropriate for rather thick films or waveguides and for
particles/waves with relatively large wavelengths.

A random rough profile $y\left( x\right) $ can be generated numerically
using some distribution function $P\left( y\right) $. The usual choice is
the Gaussian distribution,%
\begin{equation}
P\left( y\right) =\frac{1}{\sqrt{2\pi}}\exp\left( -y^{2}/2\right) ,
\label{g1}
\end{equation}
(see Ref. \cite{q2} and references therein). The simple distribution $%
P\left( y\right) $ of the type $\left( \ref{g1}\right) $ leads to an
uncorrelated roughness, $\zeta\left( x\right) \propto\delta\left( x\right) $
(white noise). To produce meaningful desirable binary correlations $%
\zeta\left( x\right) $, 
\begin{equation}
\zeta\left( x\right) =\left\langle y\left( x^{\prime}\right) y\left(
x^{\prime}+x\right) \right\rangle _{x^{\prime}}\equiv\frac{1}{L}\int y\left(
x^{\prime}\right) y\left( x^{\prime}+x\right) dx^{\prime},  \label{g2}
\end{equation}
one requires a more complicated distribution $P\left[ y\left( x\right) %
\right] $ than the straightforward distribution $\left( \ref{g1}\right) $
which is embedded in the generators of random numbers.

The first step is discretizing the surface into a large number segments, 
\begin{equation}
y\left( x\right) \rightarrow y_{i}=y\left( x_{i}\right) ,i=1,2,...,N
\label{gg2}
\end{equation}
and, if necessary, smoothing the resulting profile after the computations
are done. One way to proceed is to generate the surface with a generalized
Gaussian probability distribution, 
\begin{equation}
P\left[ \overrightarrow{y}\right] =C\exp\left( -\frac{1}{2}\overrightarrow {y%
}\cdot\widehat{G}\overrightarrow{y}\right) ,\,\overrightarrow{y}=\left(
y_{1},y_{2},...,y_{N}\right) ,  \label{g3}
\end{equation}
with some matrix $\widehat{G}$. The choice of $\widehat{G}$ in $\left( \ref%
{g3}\right) $ should provide the desired binary correlation function of
surface roughness 
\begin{equation}
\zeta\left( x\right) \rightarrow\zeta_{ik}=\zeta\left( i-k\right)
=\left\langle y_{i}y_{k}\right\rangle =\int y_{i}y_{k}P\left[ 
\overrightarrow {y}\right] d\overrightarrow{y}.  \label{g4}
\end{equation}
Here $C$ is the normalization constant defined by the equation%
\begin{equation}
1=C\int\exp\left( -\frac{1}{2}\overrightarrow{y}\cdot\widehat{G}%
\overrightarrow{y}\right) d\overrightarrow{y}.  \label{g5}
\end{equation}

If one rotates the vector $\overrightarrow{y}$,%
\begin{equation}
\overrightarrow{y}=\widehat{A}\overrightarrow{g},\,\overrightarrow{g}=%
\widehat{A}^{-1}\overrightarrow{y},  \label{g6}
\end{equation}
in such a way as to diagonalize the quadratic form $\overrightarrow{y}\cdot%
\widehat{G}\overrightarrow{y}$,%
\begin{align}
-\frac{1}{2}\overrightarrow{y}\cdot\widehat{G}\overrightarrow{y} & =-\frac{1%
}{2}\widehat{A}\overrightarrow{g}\cdot\widehat{G}\widehat {A}\overrightarrow{%
g}=-\frac{1}{2}\overrightarrow{g}\cdot\widehat{A^{T}}\widehat{G}\widehat{A}%
\overrightarrow{g},  \label{g7} \\
\widehat{A^{T}}\widehat{G}\widehat{A} & =\widehat{I}\equiv\delta_{ik},\,
\label{g8}
\end{align}
the probability distribution $\left( \ref{g3}\right) $ (including the
Jacobian) becomes%
\begin{equation}
P\left[ \overrightarrow{y}\right] d\overrightarrow{y}\rightarrow P\left[ 
\overrightarrow{g}\right] d\overrightarrow{g}=\frac{1}{\left( 2\pi\right)
^{N/2}}\exp\left( -\frac{1}{2}\sum_{i=1}^{N}g_{i}^{2}\right) d%
\overrightarrow{g}  \label{g9}
\end{equation}
meaning that all $g_{i}$ are statistically independent,%
\begin{equation}
\left\langle g_{i}g_{k}\right\rangle =\int g_{i}g_{k}P\left[ 
\overrightarrow {g}\right] d\overrightarrow{g}=\delta_{ik}\text{.}
\label{g10}
\end{equation}
The coefficient in Eq. $\left( \ref{g9}\right) $, together with the
transformation Jacobian, gives the normalization coefficient $C$ in Eqs. $%
\left( \ref{g3}\right) ,\left( \ref{g5}\right) $. Then the roughness
correlation function $\widehat{\zeta}=\left\langle y_{i}y_{k}\right\rangle $
acquires the form%
\begin{align}
\widehat{\zeta} & =\int y_{i}y_{k}P\left[ \overrightarrow{y}\right] d%
\overrightarrow{y}=\int A_{il}g_{l}A_{km}g_{m}P\left[ \overrightarrow {g}%
\right] d\overrightarrow{g}  \label{g11} \\
& =A_{il}A_{km}\delta_{lm}=\left( G^{-1}\right) _{ik}.  \notag
\end{align}
(the last equation is based on Eq. $\left( \ref{g8}\right) $). Therefore,
numerically the problem requires inverting the "desirable" matrix $\widehat{%
\zeta}$, $\widehat{G}=\widehat{\zeta}^{-1}$, and computing the rotation
matrix $\widehat{A}$. For real symmetric matrices, the rotation matrix $%
\widehat{A}$, according to Eq. $\left( \ref{g11}\right) $, is%
\begin{equation}
\widehat{A}=\widehat{\zeta}^{1/2}\text{.}  \label{g12}
\end{equation}

Summarizing, generating a random rough surface with a desirable correlation
function of surface roughness $\zeta\left( x\right) $ $\left( \ref{g2}%
\right) $ reduces to generating a set of random uncorrelated numbers $%
\overrightarrow{g}$ for a simple Gaussian distribution $\left( \ref{g1}%
\right) $ and rotating this vector using the rotation operator $\widehat{A}$ 
$\left( \ref{g12}\right) $. Computationally, this is a straightforward task.
The only limitations on the surface size, as measured in terms of step sizes 
$b=\Delta x=x_{i+1}-x_{i}$, are computational resources required to perform
the operation $\left( \ref{g12}\right) $ for large matrices $\widehat{\zeta}$%
. Obviously, this limitation is much more important for two dimensional ($%
\mathrm{2D}$) surfaces than for one dimensional ($\mathrm{1D}$) ones: in
addition to a size explosion in the $\mathrm{2D}$ case, the matrices for the 
$\mathrm{2D}$ surfaces loose their almost diagonal structure even for very
steep correlation functions.

The above procedure is straightforward in $\mathrm{1D}$. Expanding it to $%
\mathrm{2D}$ surfaces can be done in one of two ways. In principle, one can
modify the procedure by designating the raw and rotated profiles $g$ and $y$
not as vectors but as $\mathrm{2D}$ arrays and considering the rotation
operator as a 4-component tensor. We preferred instead to make a flat file
out of the $\mathrm{2D}$ surface profile and redefine the surface correlator
using this flat file. After the rotation, the points of the newly created
flat file were projected back onto the surface grid.

There is a certain ambiguity in the computation of the averages $\zeta
_{ik}=\zeta\left( i-k\right) =\left\langle y_{i}y_{k}\right\rangle $ in
samples of finite size (finite $N$). In a $\mathrm{1D}$ case, one cannot
extend evaluation of $\zeta\left( s\right) $ beyond $s=N/2$ without loosing
accuracy even if one introduces a periodic boundary condition. The same is
true when extracting the correlation function $\zeta^{\exp}\left( s\right) $
from the scanning microscopy data on the surface profile.

In a $2D$ system the loss of data points is worse. If the sample is large
enough, one can limit oneself to using $N/4$ (one quadrant of the surface)
for a straightforward calculation of the correlator up to the distances $%
\sqrt{N}/2$. If the sample size $L=\sqrt{N}$ is an issue, which is usually
the case since the required processing power is determined by $N$ and not $L$%
, one can extend the computation to approximately $N/2$ points but should
take special care to avoid double-counting of the correlations.

This technique allowed us to generate a rough surface with an arbitrary
correlation function of roughness. In numerical examples below we reproduce
three most popular types of the correlators, namely, the Gaussian,%
\begin{equation}
\varphi_{G}=\exp\left( -x^{2}/2R^{2}\right) ,  \label{g55}
\end{equation}
exponential,%
\begin{equation}
\varphi_{E}=\exp\left( -x/R\right) ,  \label{e1}
\end{equation}
and power law%
\begin{equation}
\varphi_{PL}=\frac{1}{\left( 1+x^{2}\right) ^{3/2}}  \label{pl1}
\end{equation}
correlation functions. The same correlation functions will be used as
fitting functions when probing the surfaces. All numerical parameters,
extracted with the help of these correlators, will carry the same indices $%
G,E,$ and $PL$. Note that our particular power law correlator $\left( \ref%
{pl1}\right) $ has an exponential power spectrum and \textrm{vice versa.}

Each physical system has its own spatial scale $l_{0}$. These scales for
different systems can differ from each other by orders of magnitude. It is
convenient to measure length parameters of each correlator in units of its
own physically meaningful scale $l_{0}$ leaving the definition of $l_{0}$ to
the underlying physical systems. We have three length parameters: the
average amplitude of surface roughness $\eta=\ell/l_{0},$ the correlation
radius of roughness $r=R/l_{0}$, and the step (grid) size $b=\Delta
s/l_{0}=x_{i+1}-x_{i}$, Eq. $\left( \ref{g9}\right) .$ Since the square of
the amplitude of surface inhomogeneities $\ell$ enters most of the physical
results as a simple scaling parameter, in all illustrations we assume,
unless mentioned otherwise, that $\eta=\ell/l_{0}=1$ and assign other values
to $\eta$ only when the physical situation requires this. Therefore, in
graphical illustrations below the amplitude of profile inhomogeneities can
be arbitrarily compressed resulting in smoother profiles and correlation
functions. The values of $b$ and $r$ are not necessarily independent: for
example, one can generate the correlation functions with various values of $%
r $ either by calculating the rotation matrix $\widehat{A}\left(
b=1,r\right) $ directly or calculating $\widehat{A}\left( b,r=1\right) $ and
then compressing or stretching the generated surface so that get the desired
value of $r$. In some situations $b$ is an independent physical parameter.
The most obvious example is the step size in the STM-like measurements.

The accuracy with which the generated surface reproduces the desirable
correlator $\zeta$ improves with the increase in the number of points $N$.
The limit of accessible values of $N$ depends on our ability to compute and
use the $N\times N$ matrix $\widehat{A}$, Eq. $\left( \ref{g12}\right) $ (in
our examples going to $N$ above five thousand was not practical with easily
available resources).

Even with a fixed large number of points $N,$ the standard deviation $\sigma$
between the desirable and generated correlators is not always the right way
to look at the quality of the generated rough surfaces. In general, the
correlation function consists of two parts: a peak area of the size $r$,
which describes the short-range correlations, and a long tail of long-range
correlations. Under the usual circumstances, the correlation function is
expected to go to zero at large distances. However, the correlation
functions for finite size samples inevitably contain long fluctuation-driven
tails. The same is true for experiments on restricted scanning areas. As a
result, $\sigma$ is determined mostly by these tails of the correlation
functions and is not very sensitive to the shape of the peak area.
Paradoxically, the larger the size of the sample the less sensitive can $%
\sigma$ be to the shape of the short-range peak and the rate of decrease of
the "real" correlation function. We will encounter this issue throughout the
paper.

On the other hand, the contributions from the peak and tail areas to
physical observables for different physical applications enter with
different weights: while some of the observables are more sensitive to the
short-range correlations from the peak area, the others require more
information and, therefore, better accuracy, in the tails. When the peak
area is more important, one should have more points inside the peak. The
number of such points is given by the ratio of the correlation radius $r$ to
the step size $b$, $r/b$. However, a large increase in the number of such
points leads to proportional decrease in the number of surface
inhomogeneities $Nb/r$ ("clusters" or "domains") which one can fit on the
generated surface with the fixed overall number of points $N$. This, in
turn, suppresses accuracy of the generated correlation tails and increases
the value of the overall $\sigma $ which is weighted more heavily towards
the tails of the correlation functions. This effect was obvious when we
looked at $\sigma $\ as a function of $r/b$. As a result, the computer
generation of rough surfaces with large correlation radii $r$ requires a
dramatic increase in the overall number of points which is difficult to
achieve. This also means that reliable computer simulations of theoretically
predicted physical effects at very large $r$ are not feasible. For example,
we are currently not able to reproduce computationally a new type of quantum
size effect in conductivity of ultrathin films, predicted in Ref. \cite{pon1}%
, by generating a thin film with rough surface with very large $r$. We will
encounter this issue later on in a slightly different context.

\begin{figure}[h!]
\centering
\includegraphics[scale =0.5]{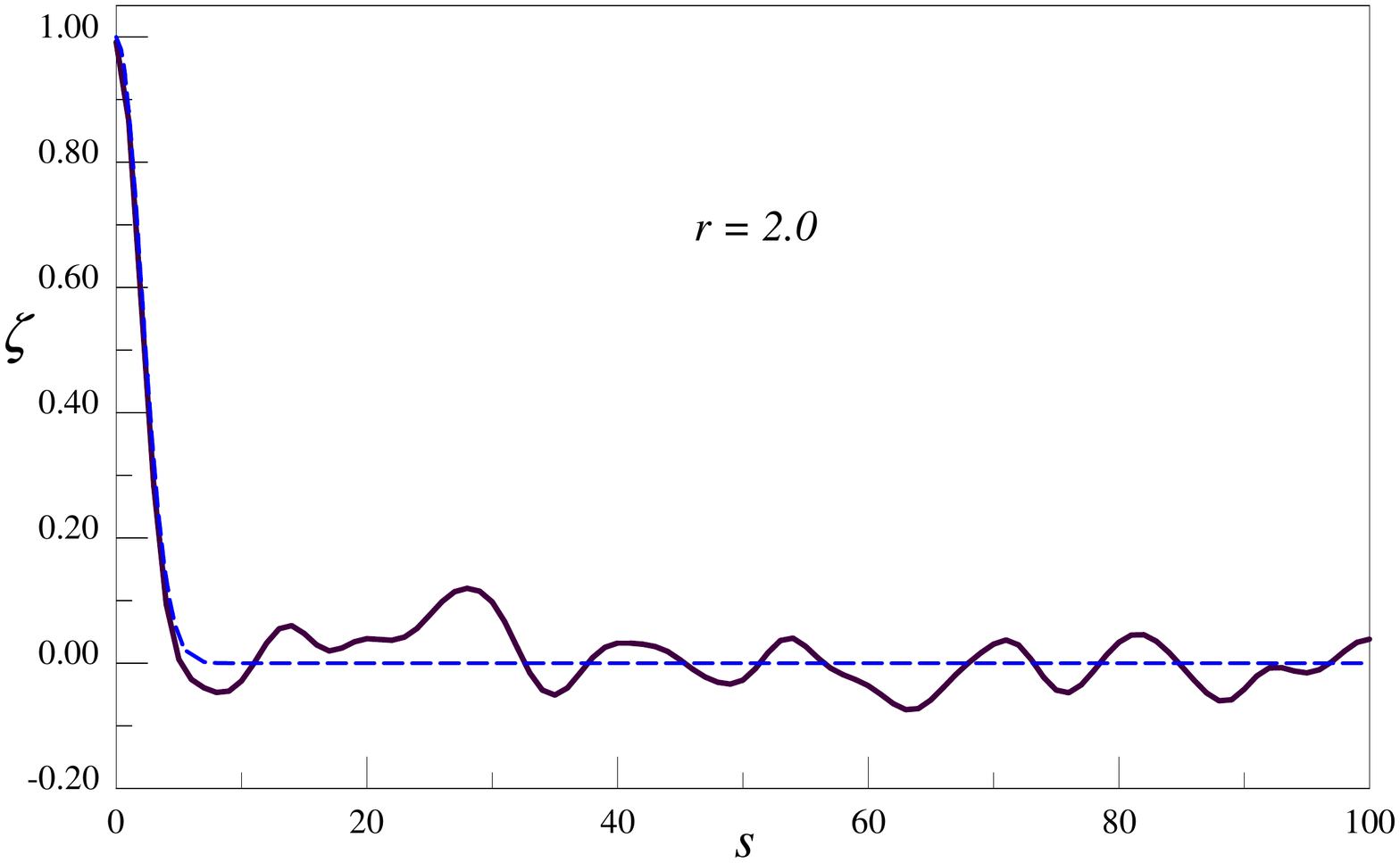}
\begin{center}
\caption{An example of the correlation function
(black solid line) for a generated $\mathrm{1D}$ surface which should
emulate a surface with Gaussian correlation of inhomogeneities $\zeta \left(
x\right) =\exp \left( -x^{2}/8\right) $ (blue dashed line). The total number
of points is 2000, the average amplitude of roughness $\eta =\ell /l_{0}=1$,
the correlation radius $r=R/l_{0}=2.$}
\end{center}
\end{figure}

\bigskip

In Figure $1$ we present the initial part of the correlation function (black
solid line) for the generated $\mathrm{1D}$ surface profile which should
emulate a surface with the Gaussian correlation of inhomogeneities $\zeta
\left( x\right) =\exp \left( -x^{2}/8\right) $ (dashed blue line). The total
number of points is $N=2000$, the average amplitude of roughness $\eta =\ell
/l_{0}=1$, the correlation radius $r=R/l_{0}=2.$ The long oscillating tail
in the correlation function reflects fluctuations.

\bigskip

\begin{figure}[htb!]
\centering

\subfigure{
   \includegraphics[scale =0.43] {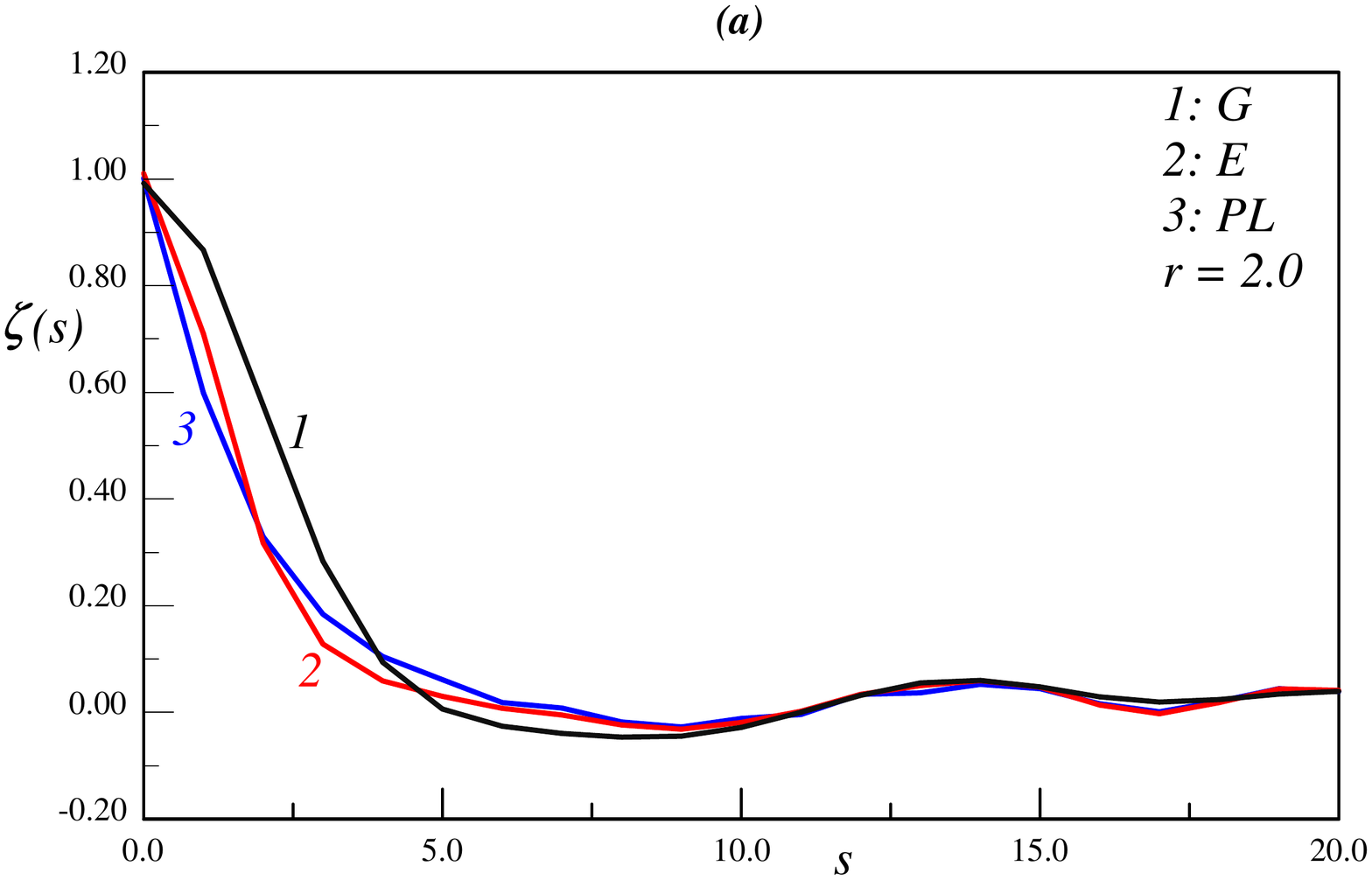}
   \label{fig:subfig1}
 }
 \subfigure{
   \includegraphics[scale =0.43] {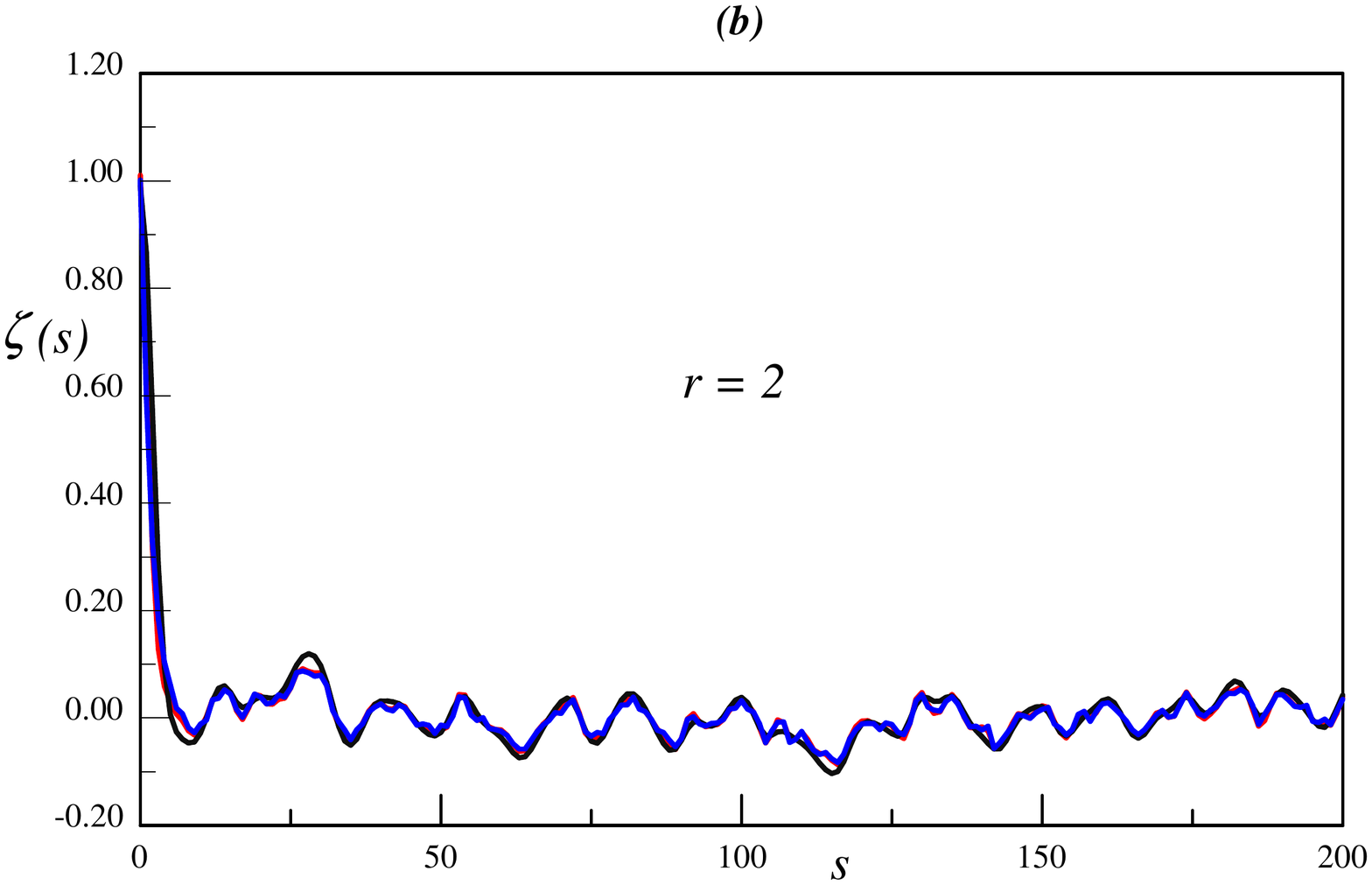}
   \label{fig:subfig2}
 }
\label{myfigure}
\caption{\textbf{\ }Correlation functions for $%
\mathrm{1D}$ generated surface profiles which should emulate the Gaussian
(black; curve 1), exponential (red; curve 2), and PL (blue; curve 3)
correlation functions. In the peak area (Figure $2a$) the differences are
very pronounced, but the fluctuation-driven tails (Figure $2b$) are almost
identical. All three computations started from the same set of $N=2000$
random numbers.}
\end{figure}

For comparison, in Figure $2$ we plotted together correlation functions
which should reproduce the Gaussian $\zeta \left( x\right) =\exp \left(
-x^{2}/8\right) $ (curve 1; black), exponential $\zeta \left( x\right) =\exp
\left( -x/4\right) $ (curve 2; red), and power law $\zeta \left( x\right)
=1/\left( 1+x^{2}/4\right) ^{3/2}$ (curve 3; blue) correlation functions
with $N=2000$, $\eta =\ell /l_{0}=1,$ and $r=R/l_{0}=2.$ In all three cases
the generation started from the same set of random numbers $\overrightarrow{g%
}$. It is clear that in the peak area (Figure $2a$) the correlations are
different, but in the tail area (Figure $2b$: the same functions $\zeta
\left( s\right) $ as in Figure $2a$ extended to $s=200$) all three curves
look the same. As a result, the quality of reproducing the desired
correlation function is the same if measured by $\sigma $ which is heavily
weighted towards the fluctuation-driven tail area.

\bigskip
 
As expected, the value of the standard deviation $\sigma $ between generated
and exact correlation functions decreases with increasing surface size $N$
as $\sqrt{2/N}$ (Figure 3).


\begin{figure}[h!]
\centering
\includegraphics[scale =0.5]{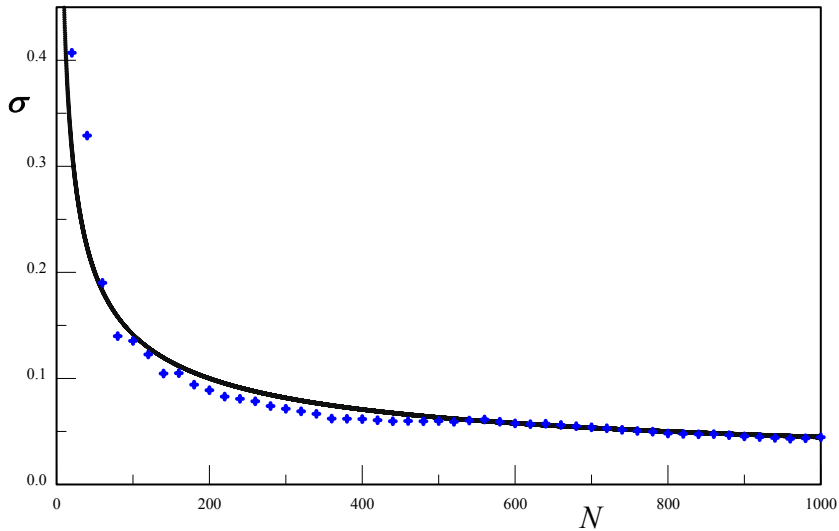}
\begin{center}
\caption{Dependence of the standard deviation $%
\sigma$ between generated and exact correlation functions on the sample size 
$N$. The solid line is $\sqrt{2/N}$. The generated roughness is supposed to
have Gaussian correlations with $r=2$.}
\end{center}
\end{figure}

Generation of $\mathrm{2D}$ roughness by this method requires more
computational resources. We were not able to routinely proceed for surfaces $%
L\times L$ with size $L$ well above $70$ when the size of the rotation
matrix $\widehat{A}$ exceeds $4900\times 4900$; computations beyond that
required special efforts. An example of a correlation function for a
generated $\mathrm{2D}$ rough surface is given in Figures 4. The roughness
correlations were supposed to emulate isotropic Gaussian correlations with $%
r=2$, $\zeta \left( s\right) =\exp \left( -s^{2}/8\right) $. Figure 4$a$
shows the $\mathrm{2D}$ correlation function $\zeta \left( x,y\right) $\ for
this surface. The anisotropy of the extracted correlation function is well
pronounced. Similar anisotropy of the extracted correlator is quite
pronounced in STM experiments as well (see, for example, Ref. \cite{munoz4}%
). After averaging over the angles, this correlation function becomes $\zeta
\left( s\right) $ in Figure 4$b$ (blue curve; the black curve gives the
emulated Gaussian correlator $\exp \left( -s^{2}/8\right) $).


\begin{figure}[htb!]
\centering

\subfigure{
   \includegraphics[scale =0.44] {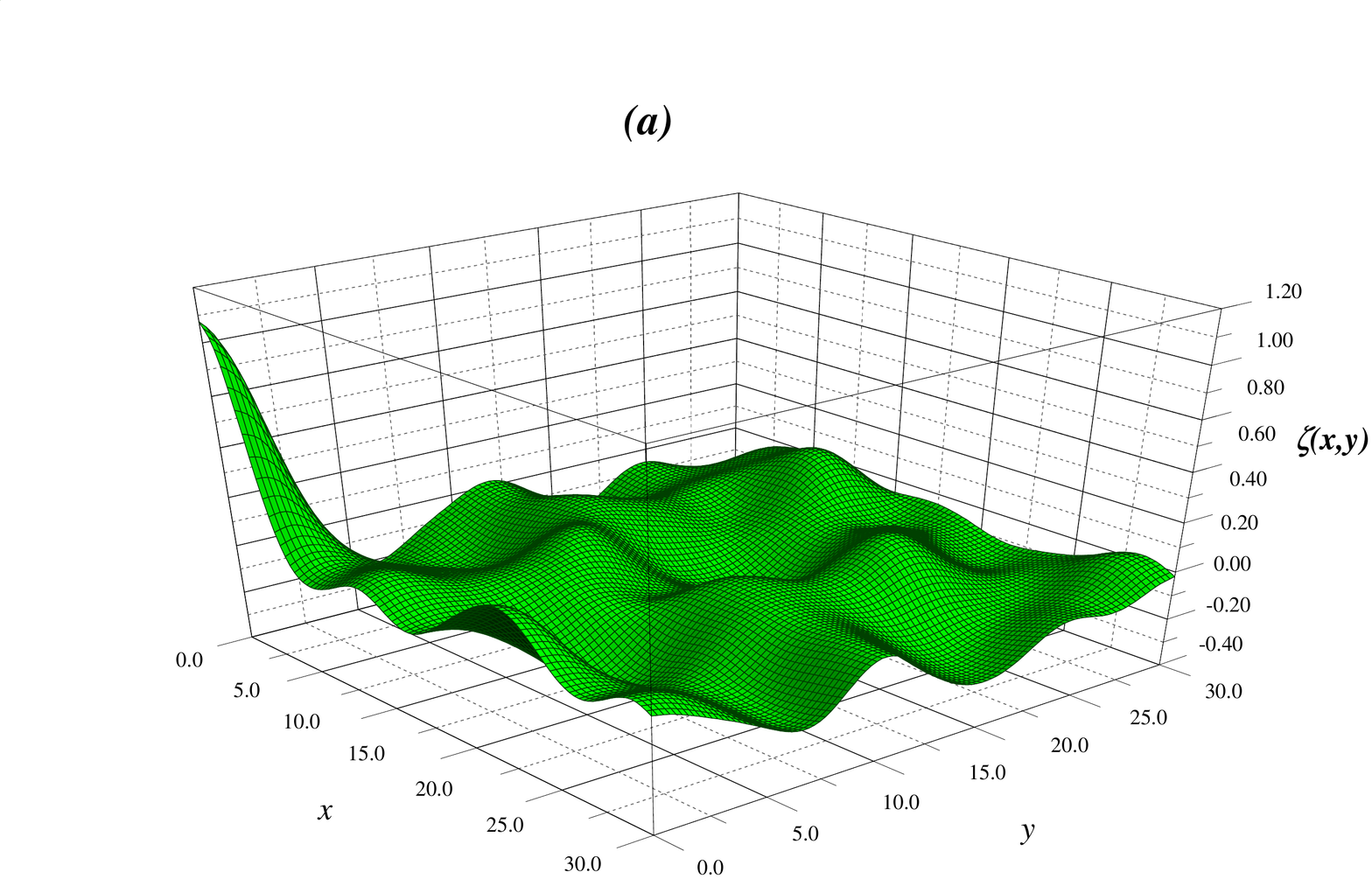}
   \label{trt1}
 }
 \subfigure{
   \includegraphics[scale =0.44] {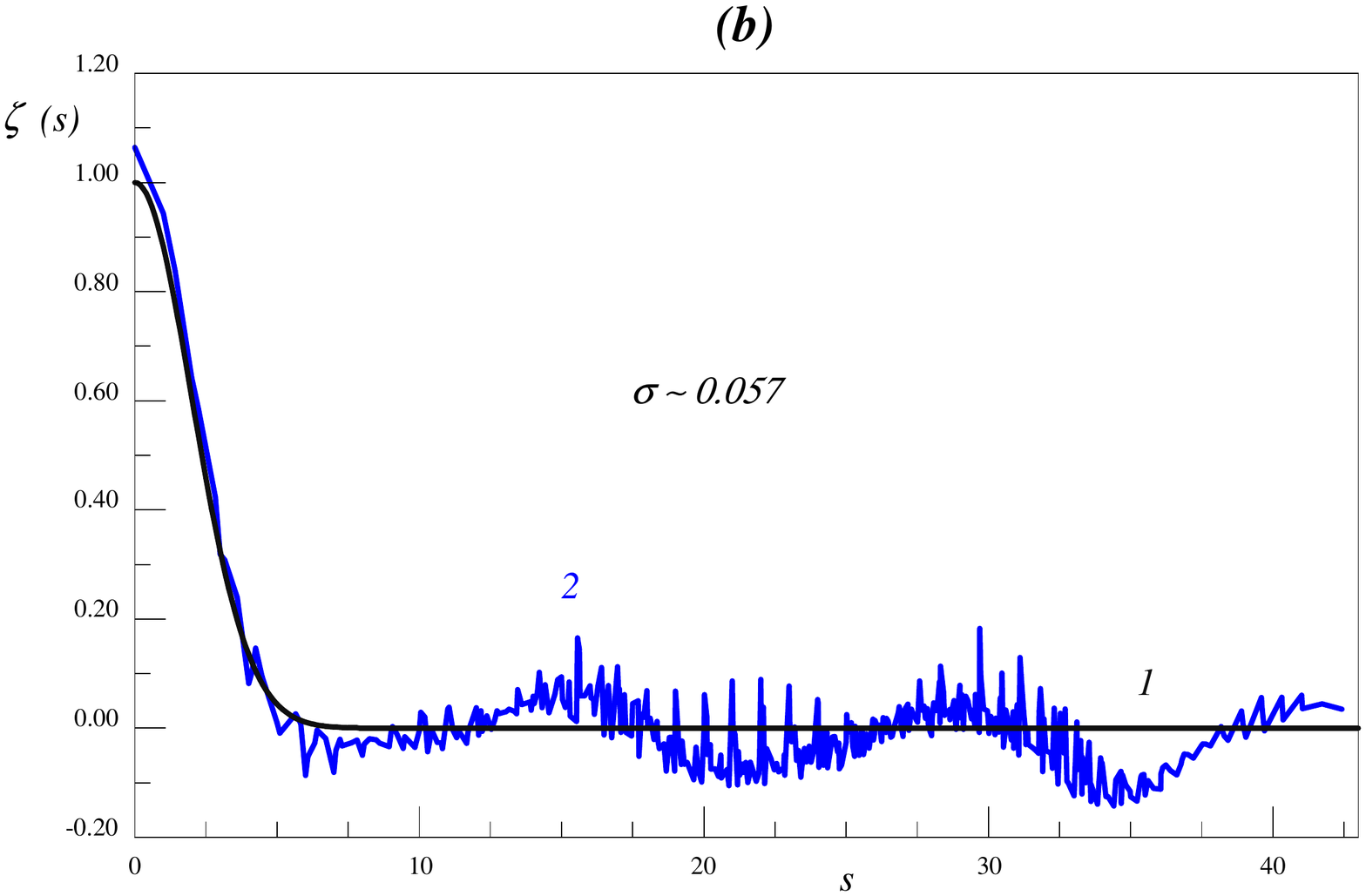}
   \label{www}
 }
\caption{An example of a $\mathrm{2D}$ rough
surface of the size $60\times 60$. The roughness emulates isotropic Gaussian
correlations with $r=2$, $\zeta \left( s\right) =\exp \left( -s^{2}/8\right) 
$ (black line 1 in Figure 4$b$). ($a$) $2D$ correlation function $\zeta
\left( x,y\right) $ ($b$) The correlation function $\zeta \left( s\right) $
after averaging over the angles (line 2; blue).
}
\end{figure}

The standard deviation between the two curves in Figure 4$b$ is surprisingly
small, $\sigma \sim 0.057$, though visually the generated correlation
function $\zeta \left( r\right) $ looks very volatile while the \textrm{2D}
function $\zeta \left( x,y\right) $ is smooth. The reason for this
volatility is quite obvious: the nearby points in $\zeta \left( r\right) $
correspond very different orientations in $\zeta \left( x,y\right) .$ With
increasing sample size $L$ the volatility actually increases because the
density of data points in $\zeta \left( r\right) $, each of which represent
different directions, goes up at large $r$. The volatility becomes so strong
that the flattened correlation function becomes unstable and practically
useless for data analysis and one should deal with the anisotropic $\zeta
\left( x,y\right) $ directly (see Section IV).

Another difficulty, which, though common to both $\mathrm{1D}$ and $\mathrm{%
2D}$ surfaces, is exacerbated in the $\mathrm{2D}$ case, concerns surfaces
with long range correlations of inhomogeneities (large $r$). The large value
of $r$ means that the surface is covered by large size inhomogeneities
(domains). The larger the value of $r$, the smaller the number of
inhomogeneities for the samples of the same linear size $L$. The
correlations of particles within each inhomogeneity are responsible for the
central peak of the radius $r$\ in the correlation function $\zeta\left(
r\right) $. However, there are noticeable non-zero correlations between the
particles from different inhomogeneities which are due not to some aligning
physical forces, but simply to geometrical factors arising from the large
size of inhomogeneities. These non-zero interdomain correlations manifest
themselves as smaller secondary peaks of the radius $r$ at positions that
correspond to the integer numbers of average distances between the domains.
If the sample is large enough to contain a very large number of such
domains, these secondary peaks are washed out. The washing out of these
peaks is determined not by the total number of data points in the sample $N$%
, which is proportional to $L$ or $L^{2}$ depending on dimensionality, but
by the ratio $N/N_{i}$ where $N_{i}$ is the number of particles in a typical
domain. If the number $N$ is not very big or the inhomogeneity clusters $%
N_{i}$ are large, these secondary peaks survive and $\zeta\left( r\right) $\
looks as if the system has an additional, larger correlation radius $R_{2}$.
The situation is worse in the \textrm{2D} case in which the number of
particles in a domain $N_{i}$ grows with $r$ as $r^{2}$. If one plots the
set of correlators for the generated surface with increasing $r$, one will
see a widening central peak and tails with more and more distinct secondary
correlation peaks. In our typical $\mathrm{1D}$\ examples with $N=2000$\ one
cannot proceed with $r$ well above $10$ without the tails loosing any
relationship to the physical forces and starting to reflect purely
geometrical interdomain correlations.\ This explains why it is so difficult
to generate rough surfaces with large $r$.

\subsection{Surfaces with discrete (integer) amplitudes of roughness}

Above we treated rough surfaces as \textrm{1D} or \textrm{2D }objects that
are described by smooth functions. This can be easily justified when the
natural physical scale of the system $l_{0}$ is much larger than the atomic
size $a$, $l_{0}\gg a$, as in our neutron example in which $l_{0}\simeq6$ $%
\mu \mathrm{m}$ (see Section IIIA below). This is a good approximation for
systems with macroscopic roughness and/or longwave particles. In the case of
electrons in ultrathin metal films the amplitudes of inhomogeneities have
atomic scale and the situation is different. The approach should depend on
whether one deals with atomically rough or atomically smooth surfaces. In
the former case, such as, for example, for amorphous films, the theoretical
description via the correlation function might still work though the
correlators should be discretized in order to account for discrete nature of
atomic-size steps in scanning measurements or computer models. In the latter
case, the rough surfaces can be understood as perfect crystal faces with
roughness introduced by randomly distributed adatoms/vacancies and steps
with kinks. If this is the case, then the roughness profile is described by
an integer number of defects of the atomic height $a$ which now becomes the
only scale of the problem. Then the use of continuous correlators for
computer modeling and STM\ data should be revisited even if one ignores the
obvious angular anisotropy. For example, the small value of the amplitude of
the correlation function $\ell$ as in experiment \cite{munoz2} might mean
that either the amplitude of roughness is indeed small or that there is
simply very few surface defects. In the latter case the meaning of the
roughness correlation radius can itself become murky.

The generation of rough surfaces with discretization of amplitudes on atomic
level cannot be done using a generic procedure of Section II: the rotation
matrix $\widehat{A}$, Eqs. $\left( \ref{g6}\right) ,\left( \ref{g12}\right)
, $ is determined solely by the desirable correlator $\widehat{\zeta }$ and
the generated surface profile $y_{i}$ does not reduce to a set of integer
numbers in terms of $a$ even if before the rotation the starting values of $%
g_{i}$ were integer. In general, the best we can do with this procedure is
to generate the set of $y_{i}$ and then round the values of $y_{i}$ to the
nearest integer number $\tilde{y}_{i}$. This, of course, changes the
correlation function. The results of this approach are illustrated in Figure
5. In this Figure we used the method of Section IIA to generate the rough
surface which emulates the exponential correlator $\zeta \left( i-k\right)
=4\exp \left( -\left\vert i-k\right\vert /2\right) $ and then rounded the
data points $y_{i}$ to the nearest integer number $\tilde{y}_{i}$. The black
curve in the Figure is the initial theoretical correlator, the red line is
the correlator $\zeta \left( \left\vert i-k\right\vert \right) =\left\langle
y_{i}y_{k}\right\rangle $ of the generated rough surface, and the blue line
is the correlator $\widetilde{\zeta }\left( \left\vert i-k\right\vert
\right) =\left\langle \tilde{y}_{i}\tilde{y}_{k}\right\rangle $ after the
discretization of the surface profile $y_{i}$ to integer numbers. As one can
see, this procedure can work at best qualitatively.


\begin{figure}[h!]
\centering
\includegraphics[scale =0.5]{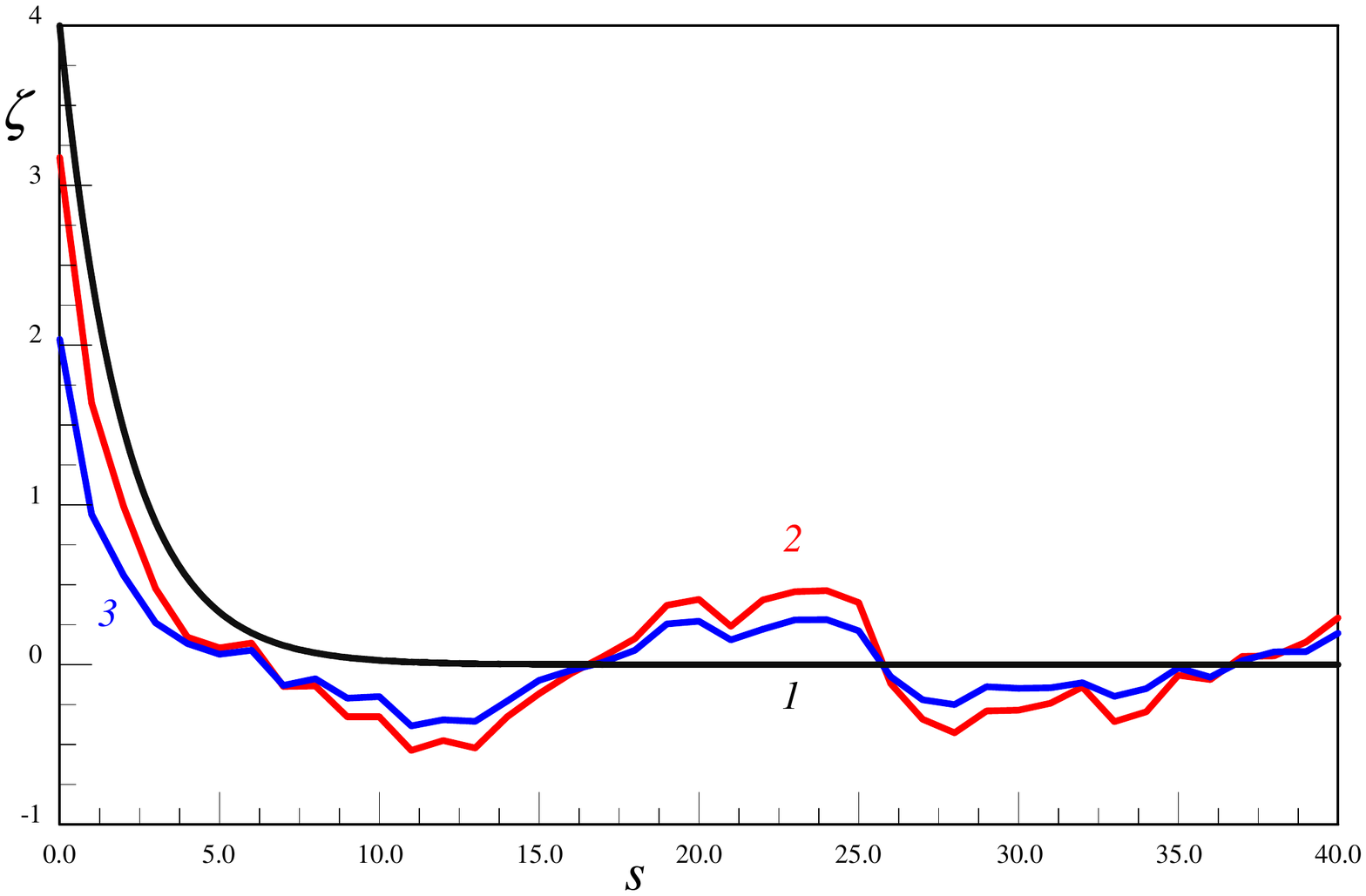}
\begin{center}
\caption{Correlation function for a generated
surface emulating $\zeta \left( i-k\right) =4\exp \left( -\left\vert
i-k\right\vert /2\right) $ (line 1; black) after rounding the profile data
points $y_{i}$ to the nearest integer number $\overline{y}_{i}$. Line 2
(red): the generated raw correlator $\zeta \left( \left\vert i-k\right\vert
\right) =\left\langle y_{i}y_{k}\right\rangle $, line 3 (blue): the
correlator for the discretized surface $\left\langle \overline{y}_{i}%
\overline{y}_{k}\right\rangle $.}
\end{center}
\end{figure}

It might be impossible to computationally emulate a random rough surface
with an integer profile $\tilde{y}_{i}$ with an \textit{arbitrary}
predetermined correlation function $\widehat{\zeta}\left( i-k\right)
=\left\langle \tilde{y}_{i}\tilde{y}_{k}\right\rangle $ except, of course,
for "classical" surfaces with very large amplitude of roughness $\ell$, $%
a\ll\ell\ll R$. However, several specific correlators can still be generated
based on spin lattice models with various Hamiltonians. This might help in
extracting the proper correlation functions from experimental data on the
surface profile based on realistic assumptions on the interaction of the
surface defects. This can also help to guess which correlation functions to
use in theoretical calculations. Needless to say, many of the lattice models
produce the correlation functions which are exponential at large distances
and have complicated, often analytically unresolved structure in the peak
area.

Unfortunately, the universe of the correlation functions which are
accessible in this way is limited by the number of known exactly solvable
lattice models, mostly in \textrm{1D}, some of which may have little
resemblance to real surfaces. It is even unclear whether there are any
restrictions on allowed forms of the correlation functions. In \textrm{2D}
even the simplest models, such as the Ising model, lead to the correlation
functions for which we do not have explicit analytical expressions making
them virtually useless for our purposes.

There are a couple of additional practical difficulties for using this
approach. First, when the correlation radius $R$ is comparable to or smaller
than the lattice constant $a$, the reliable extraction of $R$ or the shape
of the peak in the correlator from either computer generated surface or STM
data still remains impossible. In the opposite case, when $R$ is large, the
computational requirements rapidly increase because of the large size
inhomogeneities (domains). The latter requires not only going to much large
sample sizes but also an increase in computing time because of a slowdown in
convergence.

The simplest example is, of course, the ferromagnetic Ising lattice $%
y_{i}=\pm 1$ where the correlation function is determined by the attractive
coupling constant $J$ in the Hamiltonian (or, what is the same, by the
Boltzmann factors $\exp \left( \pm 2J/kT\right) $). In the \textrm{1D} case
the correlation function is exponential,%
\begin{equation}
\zeta _{E}\left( x\right) =\eta ^{2}\exp \left( -x/r\right) ,\ r=\frac{1}{2}%
\exp \left( 2J/kT\right) .  \label{a1}
\end{equation}%
The correlation function for the \textrm{2D} Ising model, though known in
principle, Ref. \cite{johnson1}, is described by a set of complicated
equations involving elliptical integrals.

\begin{figure}[htb!]
\centering
\includegraphics[scale =0.5]{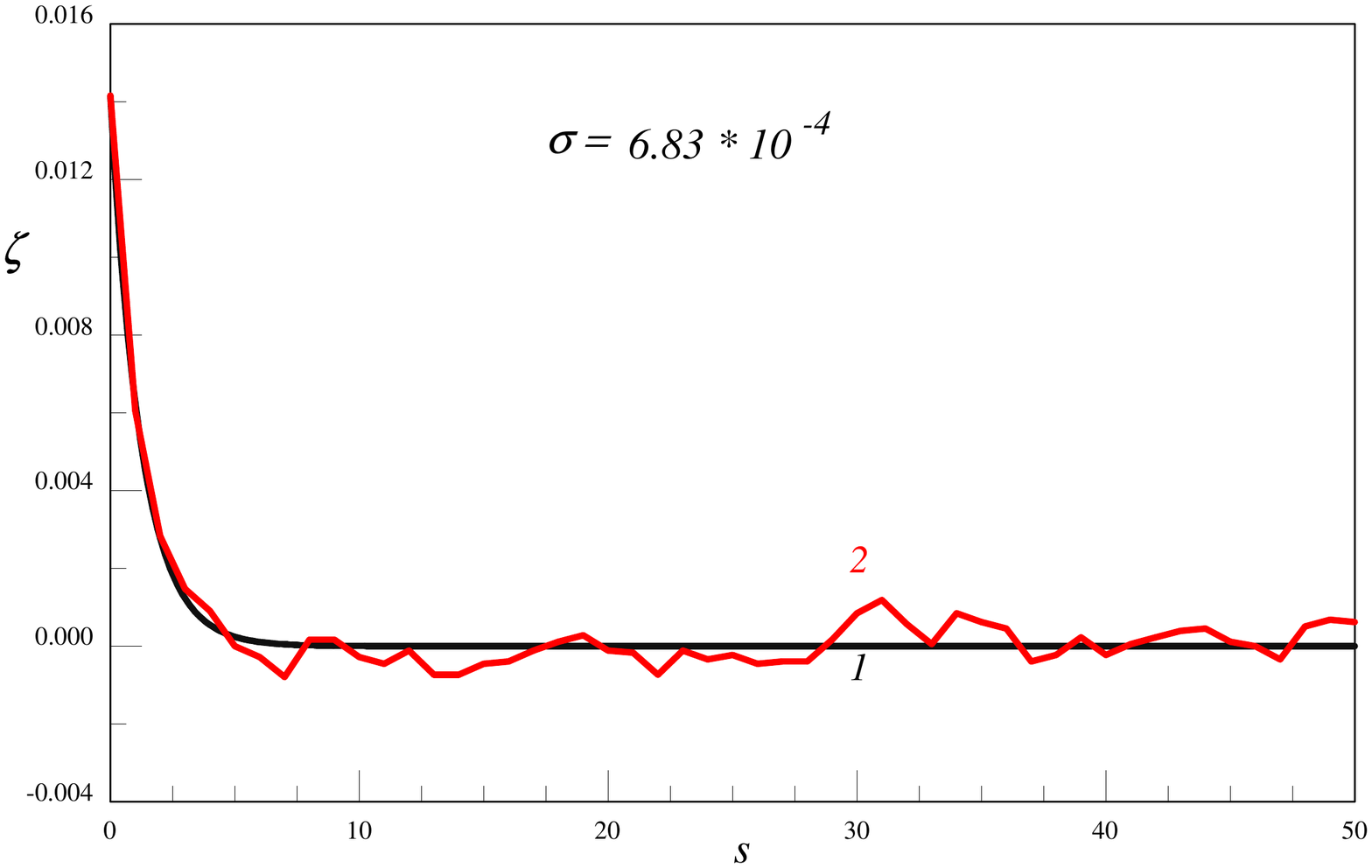}
\begin{center}
\caption{ An example of the correlation function for
a $\mathrm{1D}$ rough surface $y\left( x\right) $ and its correlation
function $\zeta \left( s\right) $ generated using the Ising model (red line
2). The parameters in the correlation function $r=1.19$ and $\eta =0.119$;
black line 1 is given by Eq. $\left( \ref{a1}\right) $ ($\sigma =6.83\times
10^{-4}$).}
\end{center}
\end{figure}

We performed Monte Carlo simulations of $\mathrm{1D}$ and $\mathrm{2D}$
rough surfaces on the basis of the Ising model. The $\mathrm{1D}$\
correlation function $\zeta \left( x\right) $ for the generated surface
profile $y\left( x\right) $ is illustrated in Figure 6. In computations we
used 1000 positions $x_{i}$ and performed $10^{6}$ Monte Carlo cycles. The
correlation function $\zeta \left( s\right) $ (red curve in Figure 6) should
emulate function $\left( \ref{a1}\right) $ with $r=R/l_{0}=1.19$ and $\eta
=\ell /l_{0}=0.119$ (black curve in the figure) as in the neutron
experiments (see below). The standard deviation between the desired and
generated correlators is $\sigma =6.83\times 10^{-4}$.

\begin{figure}[htb!]
\centering

\subfigure{
   \includegraphics[scale =0.48] {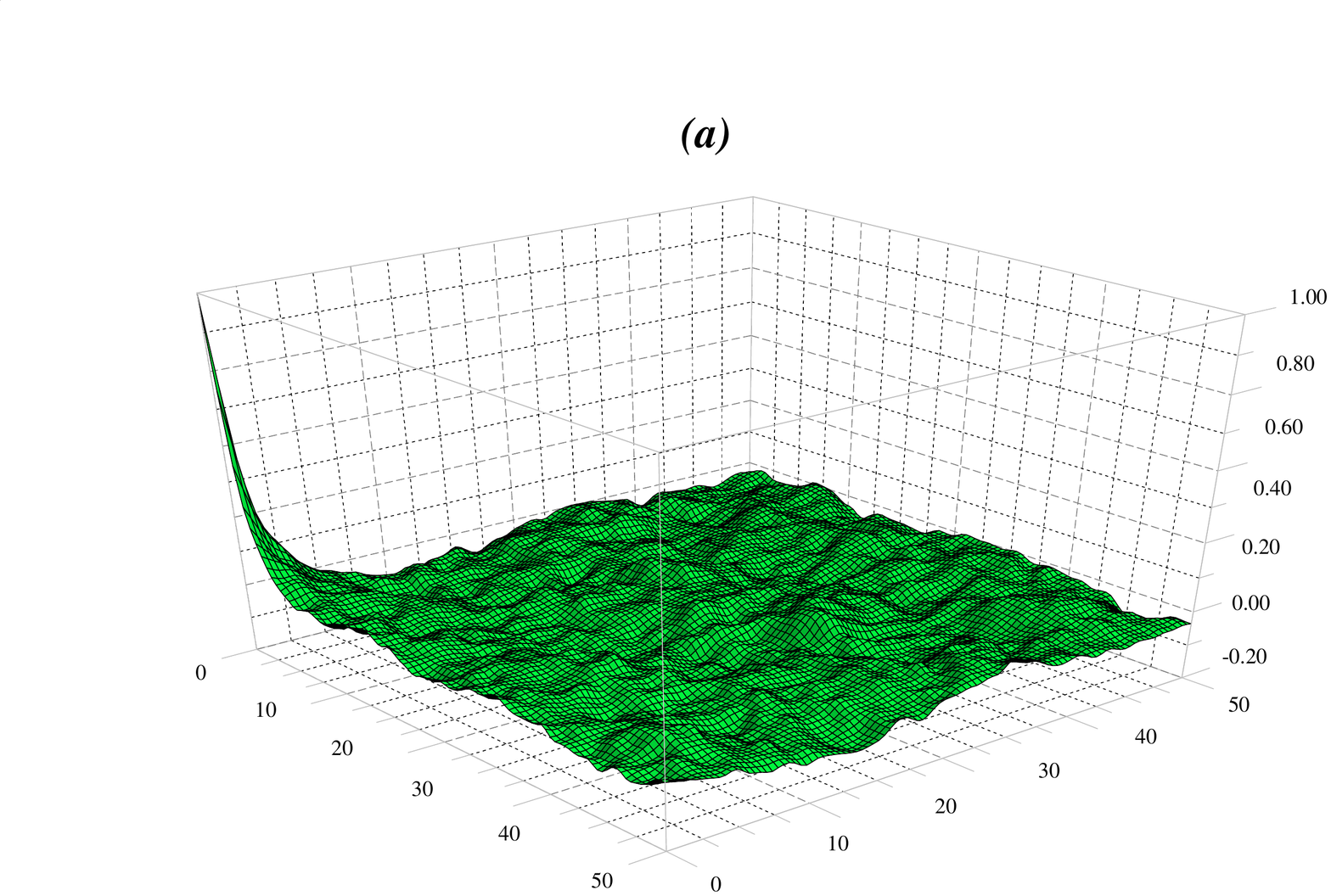}
   \label{frffe}
 }
 \subfigure{
   \includegraphics[scale =0.48] {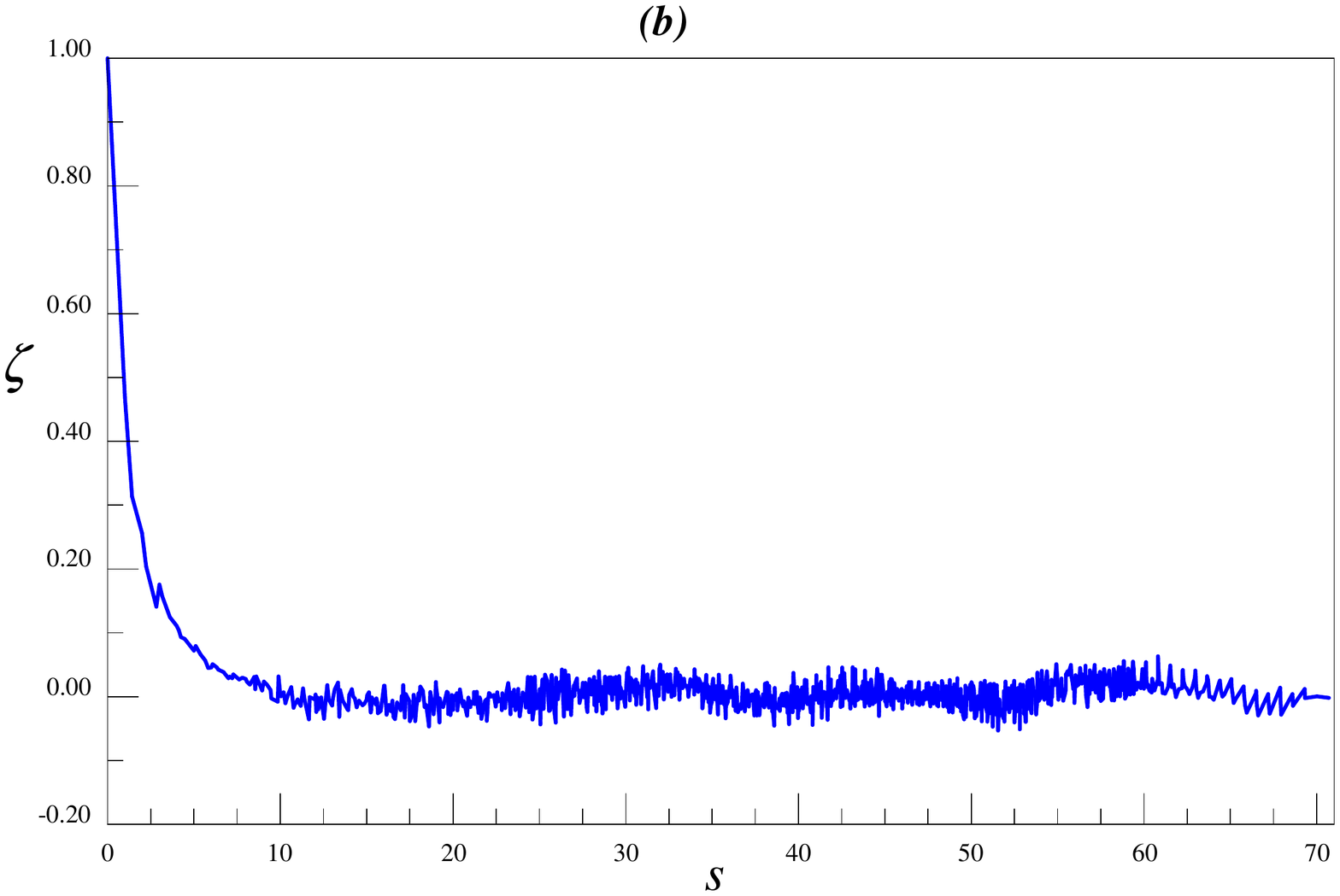}
   \label{qsa}
 }
\caption{An example of a $\mathrm{2D}$\ random
rough surface generated using the Ising model at $T=1.2T_{c}$: $a)$ \textrm{%
2D }correlation function $b)$ correlation function after averaging over the
angles.}
\end{figure}

Figures 7 illustrate the correlation function for the the surface profile
generated using the $\mathrm{2D}$ Ising model at relatively high
temperatures, $T/T_{c}=1.2$. At this temperature the relaxation (and
computation) times are not very long, domains are small, and the energy
equilibrates. On the other hand, the correlation radius already starts to
grow and the correlation function should start exhibiting deviations from
the pure exponential form. The surface area was $101\times 101$ and we
performed $10^{6}$ Metropolis cycles. Figure 7$a$ shows the \textrm{2D}
correlation function for this surface and Figure 7$b$ gives the same
correlation function after averaging over the angles. Since the sample size
is substantially larger than in Section IIA, the fluctuations are smaller.
For the same reason, the volatility of the flattened correlation function is
actually higher. A more detailed analysis is presented in Section IV. At
high temperatures this approach to generating rough surfaces seems to be
better than the one from Section IIA since it allows working with larger
samples without involving more computation resources. The situation inverses
closer to the transition and below when the relaxation times increase.

\section{Physical applications}

\subsection{\textrm{1D }applications: ultracold neutrons in a rough waveguide%
}

Experimental observation of quantization of motion of ultracold neutrons by
gravitational field \cite{neutron1} was one of the most interesting recent
achievements in neutron physics. This is a significant breakthrough in a
field with a relatively long theoretical and experimental history going back
at least into the late 1960-s; for a review and a list of publications in
the field see Ref. \cite{nesv2}. The discrete quantum states for neutrons in
the Earth gravitational field have extremely low energies with the scale of
1 \textrm{peV}. Though the quantization of motion by a linear field such as
gravity is not new by itself \cite{breit1} and has been already encountered
experimentally in a low-temperature context \cite{freed1}, the experimental
access to a spectrum of discrete energy states in such a low energy range
opens the way for using ultracold neutrons as a very sensitive probe for
extremely weak fundamental forces \cite{nesv2,pdg1,prot1,grenoble1}.

Currently, the experimental resolution of gravitational states is achieved
by sending a horizontal beam of ultracold neutrons between two horizontal
mirrors. The top mirror, the "ceiling", is intentionally made rough, while
the bottom one, the "floor", is nearly ideal (the quality of this mirror is
such that it can ensure thousands of almost specular consecutive reflections 
\cite{mirror}). The mirrors are reflective only when the normal component of
velocity is below a certain threshold; neutrons with velocities above this
threshold are absorbed by the mirrors' material. The beam of ultracold
neutrons entering this wave guide contains neutrons with a horizontal
velocity noticeably higher than this threshold and a much smaller residual
vertical component. The scattering of neutrons by the rough ceiling turns
the velocity vector thus increasing its vertical component and leading,
eventually, to absorption of the scattered neutrons. The quantization of the
vertical motion of neutrons by the Earth gravity field corresponds to the
quantization of the amplitude of bounces of neutrons from the floor mirror.
In quantum language, the turning of the velocity is equivalent to
scattering-driven transitions of neutrons into the higher quantum states.
Only the neutrons in the lowest gravitational states with the lowest kinetic
energy of vertical motion and, therefore, the lowest amplitudes of bounces,
which could not reach the rough ceiling, continue bouncing unimpeded along
the floor mirror and are counted by an exit neutron counter.

Recently we demonstrated that the results of such experiments strongly
depend on the correlation properties of roughness on the ceiling mirror \cite%
{mauricio1}. The experiments are unique in a sense that the roughness of the
ceiling mirror is created artificially. In this application, the spatial
scale $l_{0}=\hbar^{2/3}\left( 2m^{2}g\right) ^{-1/3}\sim5.871$ $\mu\mathrm{m%
}$ is the size of the lowest quantum state in the \emph{infinite}
gravitational trap (open geometry without the ceiling); $m$ is the neutron
mass. Since this length scale $l_{0}$ is relatively large, it is possible to
create the random roughness with optimized correlation properties by
computationally generating the required pattern and transferring it onto the
real surface. In earlier experiments the roughness was neither optimized nor
properly measured. The observation of the surface profile under the
microscope yielded the average distance between the nearby maxima and the
height of the peaks about 1.19 and 0.119 in units of $l_{0}$. These numbers
were accepted as the correlation radius $R$ and the amplitude $\ell$ of
surface roughness, $r=R/l_{0}=1.19$ and $\eta=\ell/l_{0}=0.119$, and the
roughness was assumed to be Gaussian. Of course, neither of these
assumptions could be justified, and the comparison of the theoretical
results to the experimental data required adjustments. If the planned new
experiments follow the suggestions of this paper, this uncertainty could be
eliminated.

An additional attraction of this system is that the geometry of the beam
experiment allows one to deal with a practically $\mathrm{1D}$ roughness
application in which the motion along the waveguide surface in the direction
perpendicular to the beam can be made irrelevant.

In Ref. \cite{mauricio1} we demonstrated that, as far as the neutron exit
count $N_{e}$ is concerned, all system parameters collapse into a single
constant $\Phi$, 
\begin{equation}
N_{e}=\sum N_{j}\left( 0\right) \exp\left( -\Phi b_{j}\left( h\right)
\right) ,  \label{ne1}
\end{equation}
where $N_{j}\left( 0\right) $ are the numbers of neutrons in the in the
quantized states $j$ that enter the rough waveguide, $h$ is the average
width of the waveguide, and $b_{j}\left( h\right) $ are the values of the
wave functions of neutrons in the gravitational states $j$ on the "ceiling".
Therefore, the problem of computationally optimizing the exit count so that
it exhibits the most pronounced step-wise dependence on the waveguide width $%
h$ reduces to increasing the value of constant $\Phi$ by manipulating the
computer-generated correlation functions.

The explicit expression for $\Phi$ via the correlation function $\psi$, Eq. $%
\left( \ref{i1}\right) $, is \cite{mauricio1} 
\begin{equation}
\Phi\left( \eta,r\right) =A\eta^{2}r\int_{0}^{1}z^{2}\psi\left( y\right) dz
\label{ne2}
\end{equation}
where the constant $A$ is determined by the size and the material of the
waveguide (in experiments \cite{neutron1, nesv2}, $A=92\cdot10^{-5}\kappa
u_{c}^{2}/\pi\chi$), and the variable $y$ in the argument of the power
spectrum of the surface roughness $\psi\left( qr\right) $ is 
\begin{equation}
y=\frac{1}{\sqrt{\chi}}\left( 1-\sqrt{1-z^{2}}\right) \sqrt{u_{c}}r.
\label{ne3}
\end{equation}
Here $u_{c}=U_{c}/mgl_{0}$, $U_{c}$ is the energy threshold for the neutron
penetration into material of the waveguide, $\chi=U_{c}/E<1$ is the ratio of
this threshold energy to the full kinetic energy of neutrons in the beam (in
past experiments $\chi\approx0.16$), and $\kappa\approx1$ differs from 1
only because of small variations in the time of flight through the waveguide.

As a result, the computation of $\Phi$, and, therefore, the expected neutron
count, for various correlation functions $\psi$ reduces, essentially, to
numerical integration in Eq. $\left( \ref{ne2}\right) $. The most important
feature here is that, because of the large value of $u_{c}$ (in experiment $%
u_{c}\sim10^{5}$), the main contribution to the integral comes from not the
whole peak in the power spectrum $\psi\left( qr\right) $, but from the
immediate vicinity of its center at $qr=0$. This means that the tails in the
computationally generated power spectrum $\psi\left( qr\right) $ are
irrelevant, but also that the only issue is to correctly reproduce the
vicinity of $\psi\left( 0\right) $. However, the closer one gets to $%
\psi\left( 0\right) $, the more important is the tail of the correlation
function $\varphi\left( x/r\right) $ at large $x$.

Our recommendation is to generate a random rough profile using the Monte
Carlo simulations on the basis of the $1D$ Ising model as described in
subsection IIB. In this particular case we prefer this method to the one
from Section IIA because it allows easier computation for a large number of
points $N$ resulting in smaller fluctuations. Another important benefit is
that the transferring the generated pattern to the real mirror is also much
simpler because all the inhomogeneities for the Ising profile have the same
constant amplitude $\pm\eta$ (Figure $6a$). In this case, the roughness
correlation function is exponential $\zeta_{E}$, Eq. $\left( \ref{a1}\right) 
$, with a simple power-law power spectrum%
\begin{equation}
\psi_{E}\left( qr\right) =1/\left( 1+q^{2}r^{2}\right) ^{3/2}  \label{ne4}
\end{equation}
which yields the following analytical expression for $\Phi$:%
\begin{equation}
\Phi_{E}\simeq\frac{1}{3}A\eta^{2}r\,_{2}F_{1}\left( \frac{3}{4},\frac{3}{2},%
\frac{7}{4},-r\sqrt{u_{c}/\chi}\right)  \label{ne5}
\end{equation}
(see \cite{mauricio1}, Eq. (38)).

Improving the experimental outcome, namely observing the well-pronounced
quantum steps in neutron count, requires the value of $\Phi$ to be\ as large
as possible. One can get the desirable value of $\Phi$ by simply
manipulating values of the correlation radius and the amplitude of surface
roughness in computer simulation. The only limitation here is that the
amplitude of the mirror roughness $\eta$ should be smaller than both the
correlation radius $r$ and the width of the waveguide $h$,%
\begin{equation}
\eta\ll r,h.  \label{ne7}
\end{equation}

When making estimates of the optimal values of $\eta$ and $r$, it is
convenient to rewrite Eq. $\left( \ref{ne5}\right) $ in the limit $%
u_{c}\rightarrow\infty$,\ 
\begin{equation}
\Phi_{E}\left( u_{c}\rightarrow\infty\right) \simeq1.38\frac{A\eta^{2}}{%
r^{1/2}}\frac{\left( 4\chi\right) ^{3/4}}{3u_{c}^{3/4}}.  \label{ne6}
\end{equation}
This equation does not have very high accuracy and for final calculations
one should still use Eq. $\left( \ref{ne5}\right) $. However, Eq. $\left( %
\ref{ne6}\right) $ highlights the dependence of $\Phi_{E}$\ on $\eta$ and $r$
in a very simple form. Since the value of $\Phi_{E}$, Eq. $\left( \ref{ne6}%
\right) $, is more sensitive to $\eta$ than to $r$, one should increase $%
\eta $ as much as possible simultaneously adjusting $r$. The limit is
imposed by the width of the waveguide which in experiment comes down to $%
h\sim2$. Therefore, the optimal waveguide should have roughness with $r=2$
and the amplitude $\eta<2$. We would not recommend to make $\eta$ much
larger than $0.2\div0.4$ - the ratio $\eta/h$ limits the accuracy of
measuring the waveguide width. Still, even this would allow to increase the
value of $\Phi$ several times in comparison to what it is assumed to be in
previous experiments with an additional benefit of ensuring a perfectly
controllable environment. The anticipated value of $\Phi_{E}$ for these
parameters is in the $43.5\div170$\ range and one should be able to see
well-pronounced quantum steps in neutron count without changing anything
else in experiment.

\subsection{\textrm{2D} applications: conductivity of ultrathin films and
surface scattering}

One of the most important applications is ballistic conductivity of
ultrathin films with rough surfaces in quantum size effect (QSE) conditions.
There are many theoretical approaches to this problem (for a short review of
earlier approaches see, for example Ref. \cite{arm2}). The equations relate
the $\mathrm{2D}$ conductivity of the films $\sigma$ to the Fourier image
(power spectrum) of the correlation function of surface roughness $%
\zeta\left( \boldsymbol{q}\right) $. Though these equations are more or less
transparent, the results, which involve inversion of large matrices, are
not. The difficulty arises because of the QSE-driven split of the $\mathrm{3D%
}$ energy spectrum $\epsilon\left( \mathbf{p}\right) $ into a large set of $%
\mathrm{2D}$ minibands $\epsilon_{j}\left( \mathbf{q}\right) $ and the
corresponding slicing of the Fermi surface. As a result, the transport
equation becomes a large set of coupled equations in the miniband index $j$.
For the purpose of this paper, namely for analysis of the correlation
functions extracted from a numerical or physical experiment, it is better to
restrict oneself to the situations in which it is possible to solve this
matrix transport equation analytically and get simple explicit expressions
for $\sigma$ via $\zeta\left( \boldsymbol{q}\right) $ (see, for example,
Ref. \cite{pon1}). We will not give here the details of the derivation and
only present these final expressions.

In the first case only the first miniband $j=1$ is occupied (ultrathin films
with very strong spatial quantization, $\hbar^{2}/mL^{2}\sim\epsilon_{F}$)
and 
\begin{equation}
\sigma=\frac{e^{2}}{3\hbar^{2}m}\tau_{1}q_{1}^{2}=\frac{2e^{2}q_{1}^{2}}{%
3\hbar^{2}m^{2}}\frac{1}{W_{11}^{\left( 0\right) }-W_{11}^{\left( 1\right) }}%
.  \label{c1}
\end{equation}
Here the lower index $1$ indicates that everything is restricted solely to
the first miniband with the Fermi momentum $q_{1}$ and $W_{11}^{\left(
0,1\right) }$ are the zeroth and first angular harmonics of the
roughness-driven scattering probabilities $W_{11}\left( \left\vert \mathbf{%
q-q}^{\prime}\right\vert \right) $,%
\begin{equation}
W_{11}\left( \left\vert \mathbf{q-q}^{\prime}\right\vert \right) =\frac{%
2\hbar}{m^{2}L^{2}}\zeta\left( \left\vert \mathbf{q-q}^{\prime }\right\vert
\right) \left( \frac{\pi}{L}\right) ^{4}  \label{c2}
\end{equation}
over the angle between the vectors $\mathbf{q}$ and $\mathbf{q}^{\prime}$
(this equation assumes that the correlation properties of both surfaces of
the film are the same and that there are no intersurface correlations).

The second analytical case is the case of small $qR$. In this limit, the
correlation function is a constant with the zero first harmonic, 
\begin{equation}
W_{jj}^{\left( 0\right) }=2W\left( qR\rightarrow0\right) ,\ W_{jj}^{\left(
1\right) }=0  \label{c3}
\end{equation}
with 
\begin{equation}
W_{jj^{\prime}}\left( qR\rightarrow0\right) =\frac{2\hbar}{m^{2}L^{2}}%
\zeta\left( qR\rightarrow0\right) \left( \frac{\pi j}{L}\right) ^{2}\left( 
\frac{\pi j^{\prime}}{L}\right) ^{2}.  \label{eee10}
\end{equation}
and 
\begin{equation}
\sigma=\frac{2e^{2}}{\hbar}\frac{\left( L/\pi\right) ^{4}}{2S\left(
S+1\right) \left( 2S+1\right) \zeta\left( qR=0\right) }\sum_{j}\left( \frac{%
Lq_{j}}{\hbar j}\right) ^{2}.  \label{ee11}
\end{equation}
where $S$ is the total number of the occupied minibands (the number of
slices of the Fermi surface by quantizing planes $p_{zj}=\pi j\hbar/L$)
which is given by the equation 
\begin{equation}
S\left( L\right) =\left\lfloor p_{F}L/\pi\hbar\right\rfloor  \label{c4}
\end{equation}

Eqs. $\left( \text{\ref{c3}}\right) -\left( \text{\ref{ee11}}\right) $
involve the power spectrum of the surface roughness $\zeta\left( q\right) $
at $q=0,$ which acquires the following form after the angular integration
(see, for example, \cite{pon1}):%
\begin{equation}
\zeta\left( q=0\right) =2\pi\int\zeta\left( \left\vert \mathbf{s}\right\vert
\right) d^{2}s=2\pi J_{0}\left( 0\right)  int \zeta\left( s\right) sds.
\label{c5}
\end{equation}

The value of $\zeta_{0}\equiv\zeta\left( q=0\right) $ is important not only
for conductivity of ultrathin films, but also for a much more general class
of problems associated with scattering longwave particle (or waves) on rough
surfaces. Scattering in longwave limit $q\rightarrow0$ is always described
by a single constant, which here is, essentially, $\zeta_{0}$. Therefore, $%
\zeta_{0}$ is one of the most important characteristics of the surface which
determines a large number of observables.

Note, that all our surface correlators $\zeta\left( s\right) $ in Ref. \cite%
{pon1} are introduced in such a way that in the longwave limit $\zeta\left(
q\rightarrow0\right) \rightarrow2\pi\ell^{2}R^{2}.$ In what follows we will
evaluate $\zeta\left( q\rightarrow0\right) $ for rough surfaces. Here one
has a choice: either to fit the correlation function extracted from the
experimental or numerical data to one of the model correlators and to get $%
2\pi\ell^{2}R^{2}$ from the best fit values of $\ell$ and $R$, or to get $%
\zeta\left( q=0\right) $ directly from the data by, for example, direct
numerical integration $\left( \text{\ref{c5}}\right) $ of the extracted
data. The accuracy with which we will be able to evaluate the value of $%
\zeta_{0}$\ will provide a much more reliable physical evaluation of the
data and techniques than the standard deviation between the extracted
correlators and the fitting functions.

\section{Probing and identifying the rough surfaces}

In this section we will analyze what kind of information one can extract
from the surfaces generated by methods of Section II. This will also give us
an insight into difficulties facing experimentalists trying to extract the
correlation function from the scanning microscopy. In this regard, the
problems facing the computational physicists and experimentalists are
roughly the same: limited sample sizes, noticeable fluctuations, large
domains, and long relaxation times. Since we know exactly how the "true"
roughness correlation function should look like when we are using the
methods of Section II, we are able to point at potential pitfalls in
extracting information from experimental and computational data when one
does not know the "true" correlation function. We will judge the quality of
the surface analysis not by the standard deviation $\sigma$ between the
extracted data and a fitting function, but by the values of the physically
important variables - $\Phi$ for the \textrm{1D} neutron problem and $%
\zeta_{0}=\zeta\left( q\rightarrow 0\right) $ for the conductivity of and
scattering from \textrm{2D} films in the longwave limit as explained in
Section III.

Our goal here is to show that the use of a fitting function of a wrong shape
can invalidate both the computations and the experiments. First, the
parameters of the correlation function $\zeta^{fit}$ obtained from the best
fit to $\zeta^{\exp}$ strongly depend on the \emph{assumption} about the
functional form (shape) of the "real" correlator. For example, the analysis
of the same STM measurements of $\zeta^{\exp}$ on the basis of the Gaussian
and exponential correlators in Ref. \cite{munoz3} provided vastly different
values of the correlation radius $R$. Since the shape of the "real"
correlator is not known \textrm{a priori}, it is almost impossible to know
what function should be fitted to $\zeta^{\exp}$ and what is the reliability
of the extracted parameters.

Note that the standard deviation $\sigma$ is supposed to be the deviation
between the extracted ("measured") correlator $\zeta^{\exp}$ and the "true"
correlation function. When the "true" correlator is unknown, as in most
experiments, what is presented as $\sigma$ is the deviation between the
extracted correlator and the models used for fitting,%
\begin{equation}
\sigma^{2}=\frac{2}{N}\sum_{j=1}^{N/2}\left(
\zeta_{j}^{\exp}-\zeta_{j}^{fit}\right) ^{2}.  \label{p1}
\end{equation}
which describes the quality of the fit and tells us nothing about
appropriateness of the fitting function.

Eq. $\left( \ref{p1}\right) $ is highly weighted towards the tails of the
correlator, especially when the correlation radius is comparable to the
probing step. When the fitting functions rapidly go to zero at large
distances, $\sigma$, Eq. $\left( \ref{p1}\right) $, does not even depend
much on the choice of the fitting function while the physical results
clearly do. When one uses the short-range fitting functions, the presence of
long fluctuation-driven tails may even emulate the presence of some
additional large correlation radius $R_{2}$ introduction of which can
noticeably decrease $\sigma$. [The second, large correlation radius was
observed, for example, in Ref. \cite{parpia1}. We do not know how to verify
the reliability of such conclusions without doing the same measurements on a
relatively large number of other surfaces, including much larger sample
sizes. This, of course, is not practical for the already difficult
experiments]. The outsize effect of these fluctuation tails often seems even
more important than the difficulties in resolving the shape of the main
maximum \cite{ogilvy2}.

The second intrinsic difficulty arises when dealing with surfaces with a
large correlation radius $r$. The large value of $r$ means that the surface
is covered by large size inhomogeneities (domains). As explained in the end
of Section IIA, the presence of a small number of large domains gives rise
to the appearance of spurious secondary peaks of the radius $r$ at positions
that correspond to the integer numbers of average distances between the
domains. These peaks reflect interdomain correlations and not physical
interactions.

In addition to analyzing the extracted correlators with the help of various
fitting functions, we will also perform the direct Fourier analysis of the
correlation data sets as it is sometimes done for experimental data. This
should, in principle, give us the full power spectrum of the correlations
which we use for direct calculation of observables. This approach allows one
to avoid the pitfall of using the fitting functions of a "wrong" shape.
However, this approach encounters difficulties of a different type. It
utilizes all the erroneous information which is contained in the
fluctuations while all the fitting function of "right" and "wrong" shapes,
which all rapidly go to zero at large distances, simply disregard the long
fluctuation-driven tails. Of course, one can always introduce the high
frequency cutoff when doing the spectral analysis, but then the physical
results become dependent on the guess for the cutoff.

We start from the $\mathrm{1D}$ case in application to our neutron problem.
Table I contains examples of three runs based on Section IIA. In each run we
generate a rough surface with a Gaussian correlation of inhomogeneities with 
$r=1.19$ and $\eta=0.119$ which is supposed to be close the real
experimental setup. The main physical parameter $\Phi$, Eq. $\left( \ref{ne2}%
\right) $, which determines the neutron count behind the waveguide, for such
roughness is equal to $\Phi=23.48$. After each numerical run, we fit the
observed correlation function with a Gaussian, $\left( \eta^{G}\right)
^{2}\exp\left( -s^{2}/2r^{G2}\right) $, exponential, $\left( \eta^{E}\right)
^{2}\exp\left( -s/r^{E}\right) $, and power law, $\left( \eta^{PL}\right)
^{2}/\left[ 1+\left( s/r^{PL}\right) ^{2}\right] ^{3/2}$, correlators and
extract the best fitting values $\eta^{G,E,PL}$ and $r^{G,E,PL}$ for the
amplitude and the correlation radius. Then we recalculate the value of $\Phi$%
, Eq. $\left( \ref{ne2}\right) $, using these fitting functions. The Table
contains values of $\Phi$ for these three types of fitting correlators, $%
\Phi^{G,E,PL}$, the standard deviations $\sigma^{G,E,PL}$ for the fittings,
and the fitting parameters $r_{fit}^{G,E,PL}$ which provide the best fits.
To save space, we do not present the fitting parameters $\eta_{fit}^{G,E,PL}$
which were all in the range $0.118\div0.123$. The Table also contains the
values $\Phi_{num}$ and $\sigma^{num}$ obtained from the Fourier analysis
with a large number of harmonics (half the number of the data points; thus a
vanishingly small value of $\sigma_{num}\sim2\times10^{-17}$). The latter
procedure is, essentially, equivalent to direct numerical integration $%
\left( \ref{ne2}\right) $ of the power spectrum of the observed correlator
with all its fluctuation-driven tails.

\begin{center}
\begin{tabular}{|c|c|c|c|c|c|}
\hline
\# & $r_{G},\ \sigma_{G}\times10^{4}$ & $r_{E},\ \sigma_{E}\times10^{4}$ & $%
r_{PL},\ \sigma_{PL}\times10^{4}$ & $\sigma_{num}\times10^{17}$ & $\Phi
_{G},\ \Phi_{E},\ \Phi_{PL},\ \Phi_{num}$ \\ \hline
1 & 1.19, 5.24 & 1.59, 5.81 & 1.44, 5.81 & 1.92 & 23.86, 18.19, 18.81, 21.96
\\ \hline
2 & 1.15, 4.49 & 1.53, 4.56 & 1.36, 4.64 & 1.83 & 23.33, 17.84, 18.65, 21.14
\\ \hline
3 & 1.25, 4.37 & 1.69, 4.40 & 1.54, 4.47 & 1.69 & 23.56, 17.26, 17.85, 20.96
\\ \hline
\end{tabular}
\end{center}

\textbf{Table I.} Three numerical runs based on Sec. IIA in application to
our neutron problem. Rough $1D$\ surfaces emulate Gaussian correlation of
inhomogeneities $\eta ^{2}\exp \left( -x^{2}/2r^{2}\right) $ with $r=1.19$
and $\eta =0.119$ as it was assumed in experiment. The expected value of the
main physical parameter $\Phi $, Eq. $\left( \ref{ne2}\right) $, is $\Phi
=23.48$. The extracted correlators were fitted with Gaussian, $\eta
_{G}^{fit}\exp \left( -s^{2}/2r_{G}^{fit}\right) $, exponential, $\eta
_{E}^{fit}\exp \left( -s/r_{E}^{fit}\right) $, and power law, $\eta
_{PL}^{fit}/\left[ 1+\left( s/r_{PL}^{fit}\right) ^{2}\right] ^{3/2}$
fitting functions. The table contains the best fitting values of $%
r_{G,E,PL}^{fit}$, together with $\sigma _{G,E,PL}$, and the recalculated
values of $\Phi _{G,E,PL}$. The columns with $\Phi _{n}$ and $\sigma _{n}$
give the values of $\Phi $ and the standard deviation when the spectral
decomposition of the data was put directly into equations without fitting.

The results are very informative. The quality of the fits $%
\sigma^{G},\sigma^{E},\sigma^{PL}$ for all three types of the fitting
functions were more or less the same, about $5\times10^{-4}$, but the
results for the physically important parameters $\Phi^{G,E,PL}$ differed
considerably, by about 25\%. In our experiment, the "true" shape of the
correlation function was known to be Gaussian and, not surprisingly, the
fitting by the Gaussian function produced the values of $\Phi$ very close to
the "true" value $23.48$. This brings us to an inevitable conclusion that
the quality of fit ($\sigma$) of measured surface correlations by some 
\textrm{ad hoc }correlator does not tell much about the quality of physical
conclusions. Note that the results for fitting by the power law and
exponential correlation functions were relatively close to each other and
very different from those for the Gaussian fit. The explanation is simple:
the Gaussian function has a much shorter tail.

Interestingly, feeding the Fourier image of experimentally observed
correlator directly into the equations without any fitting does not do much
to improve the quality of conclusions. The reason is simple: with this
approach we are using too much information about the long distance
correlations, which are determined by the fluctuations and not by any
physical forces. Still, this approach for our $\mathrm{1D}$ physical problem
works a little bit better than making a wrong guess on the shape of the
correlation function.

The results for our $\mathrm{2D}$ problem on conductivity of films are
different because of different dimensionality and smaller linear\ sizes of
our samples. The corresponding results are given in Table II. Here we were
generating the Gaussian rough surface with the correlation function $%
\zeta\left( \left\vert \mathbf{s}\right\vert \right) =\exp\left( -\left\vert 
\mathbf{s}\right\vert ^{2}/8\right) $ $\left( \emph{i.e.}\text{, }%
\eta=1,r=2\right) $ for which the theoretical value of $\zeta_{0}\equiv
\zeta\left( q=0\right) =8\pi\approx25.13$. The sample size was $61\times61$
points. The Table contains the results extracted from the best fit of the
extracted correlator to the Gaussian, exponential, and power law functions.
The quality of the fits (the values of $\sigma$) here is worse than in the $%
\mathrm{1D}$ case above though the overall number of the data points is
larger (3600 \textrm{vs.} 2000 points): the linear size of the sample is
noticeably smaller while the correlation radius is bigger. The Table
provides the values of the extracted fitting parameters $\eta$ and $r$,
values of $\sigma$, and, most importantly, the corresponding values of the
physical observable $\zeta_{0}$.

\begin{center}
\begin{tabular}{|c|c|c|c|c|}
\hline
\# & $\eta_{G1},\ r_{G1},\zeta_{0}^{G1},\sigma\times10^{2}$ & $\eta _{G2},\
r_{G2},\zeta_{0}^{G2},\sigma\times10^{2}$ & $\eta_{E},\
r_{E},\zeta_{0}^{E},\sigma\times10^{2}$ & $\eta_{PL},\
r_{PL},\zeta_{0}^{PL},\sigma\times10^{2}$ \\ \hline
1 & 1.04, 1.97, 25.3, 5.7 & 1.04, 1.95, 25.89, 9.5 & 1.14, 2.04, 33.81, 6.2
& 1.08, 2.45, 44.03, 5.9 \\ \hline
2 & 1.10, 1.80, 24.37, 6.5 & 1.10, 1.78, 27.91, 8.5 & 1.20, 1.76, 27.91, 7.4
& 1.14, 2.15, 37.75, 7.2 \\ \hline
3 & 0.90, 1.84, 24.37, 4.1 & 0.91, 1.82, 17.09, 6.0 & 0.98, 2.05, 25.17, 4.3
& 0.94, 2.40 , 31.04, 4.1 \\ \hline
4 & 1.00, 1.98, 25.0, 1.9 & 1.00, 1.97, 25.0, 2.4 & 1.10, 2.11, 33.8, 2.9 & 
1.05, 2.49, 42.9, 2.4 \\ \hline
\end{tabular}
\end{center}

\textbf{Table II.} The same as in Table I for generated $2D$ rough Gaussian
surfaces with $r=2$ and $\eta=1$. The expected value of the main physical
observable $\zeta\left( q=0\right) =8\pi.$ The table contains the extracted
fitting parameters $\eta_{G,E,PL}^{fit}$ and $r_{G,E,PL}^{fit}$, together
with $\sigma_{G,E,PL}$, and the recalculated values of $\zeta_{0}$. The
Gaussian fit was done independently for the \textrm{1D} correlation function 
$\zeta\left( \left\vert \mathbf{s}\right\vert \right) $ (index 1) and the 
\textrm{2D} correlation function $\zeta\left( \mathbf{s}\right) $ (index 2).
The fourth row gives the results for the correlation function averaged over
10 independent runs.

In this Table, the first three rows represent three different numerical
runs. The fourth row provides the results of averaging of the data extracted
from 10 numerical runs (in experiment, this is equivalent to averaging the
data extracted from 10 different pieces of the same surface).

As it was explained in Section II, we have two options when dealing with $%
\mathrm{2D}$ correlations which, because of fluctuations, always exhibit
anisotropy even when the underlying true physical correlator is isotropic.
We can either deal with an anisotropic $\mathrm{2D}$ correlator $\zeta\left( 
\mathbf{s}\right) $ and fit it to $\mathrm{2D}$ fitting functions, or
flatten the observed $\mathrm{2D}$ correlator to a $\mathrm{1D}$ function $%
\zeta\left( \left\vert \mathbf{s}\right\vert \right) $ by averaging away the
anisotropy. The drawback of the latter procedure is that the resulting $%
\mathrm{1D}$ correlator, as explained above, exhibits increasing volatility
at large distances. This volatility is not very important when using our
simple fitting functions which vanish at large distances anyway, but makes
it impossible to perform the Fourier analysis of unfitted experimental data
and feed the results directly into equations as it was done for our $\mathrm{%
1D}$ neutron problem. In this case, the results were simply unstable.

We use both options when fitting using the Gaussian correlator. In the first
column of the Table we present results obtained from the flat ($\mathrm{1D}$%
) file $\zeta\left( \left\vert \mathbf{s}\right\vert \right) $. The second
column gives the results of fitting $\zeta\left( \mathbf{s}\right) $ by a $%
\mathrm{2D}$ Gaussian function. For exponential and power law correlators in
columns 3 and 4, we use only the flattened file $\zeta\left( \left\vert 
\mathbf{s}\right\vert \right) $.

Feeding the results of the spectral analysis of extracted raw correlation
function directly into the equations (the last column), as it was done for
the neutrons, does not work at all - the results for $\zeta\left( q=0\right) 
$ are unstable because of anisotropic fluctuations. Possibly, this procedure
might have been used if we would had been able to increase the sample size.
We are not able to check this because time and memory requirements are
increasing as $L^{4}$ with an increase in the linear size $L$. However, it
is not clear whether increasing the linear size $L$ would have been of much
help: the amplitude of fluctuations would have indeed gone down, but the
length of the fluctuation tails would have increased. This procedure was
giving stable, but still not very good results, when we use the correlation
function averaged over 10 runs (the last row). Here the value of $%
\zeta_{0}^{n1}$ obtained from $\zeta\left( \left\vert \mathbf{s}\right\vert
\right) $ is $18.79$, and $\zeta_{0}^{n2}$ obtained from the $\mathrm{2D}$
Fourier analysis of $\zeta\left( \mathbf{s}\right) $ is $17.6$. Even these
two numbers are much worse than those obtained with the help of fitting
functions. However, averaging the correlation function over several runs
(or, in experiment, over several parts of the rough surface \cite%
{munoz2,munoz4}) to decrease the anisotropic fluctuations is an inherently
dangerous procedure. It can work well if one knows beforehand that the
"true" correlator is a simple slowly decreasing function. If, for example,
the correlation function contains an oscillating tail, this averaging could
destroy important physical information. The same uncertainty does not allow
one to simply cut off the fluctuation tails.

As one can see from Table II, here, as in the case of neutrons, the values
of $\sigma$ are similar for all fitting functions, but only the fitting
function with the correct (Gaussian shape) yield good results for the
physical observables (in this case, $\zeta_{0}$). In contrast to the neutron
problem, the accuracy of the results for conductivity obtained with the help
of wrong fitting functions, which is not good by itself, is nevertheless
preferable to putting the Fourier image of the raw data directly into the
equations.

The next two tables provide the similar data analysis for surfaces generated
using the Ising model. Table III presents the results of five runs for the
application of the $\mathrm{1D}$ Ising generator to the neutron problem. The
data in the columns are arranged similarly to Table I. The parameters of the
"true" correlation function are the same, $r=1.19$ and $\eta=0.119$.
However, since the Ising model corresponds to the exponential correlation
function, Eq. $\left( \ref{a1}\right) $, and not to the Gaussian correlator
as in Table I, the true value of parameter $\Phi$, Eq. $\left( \ref{ne2}%
\right) $, is now $\Phi_{E}^{th}=19.5$ (with the same values of $r$ and $%
\eta $ $\Phi_{G}^{th}=23.7$ and $\Phi_{PL}^{th}=20.4$). Since the simulation
is based on the Ising model with spins $\pm1$, the extracted average
amplitudes of roughness differ from $\eta=0.119$ by less than 1\% for all
fitting functions and there is no need to present the values of $%
\eta_{E,G,PL}$. The size of the sample was $N=1000$ and we performed $10^{6}$
Metropolis cycles. Of course, the fit using the exponential correlator
provides the best values for $\Phi$. Of the other two fits, it is not at all
clear why in this case the power law fit works much better than the Gaussian
one. The last column in the Table gives the values of $\Phi_{n}$ which
obtained by direct spectral analysis with $N/2$ harmonics of the raw
correlation data without any fitting. These data display the worst agreement
with $\Phi_{E}^{th}=19.5$\ while the value of $\sigma_{n}$ is by 13 orders
of magnitude better than $\sigma$ for any of our fitting functions. The
explanation is the same as before: the full set of raw data is dominated by
the long correlation tails which come from the fluctuations.

\begin{center}
\begin{tabular}{|c|c|c|c|c|c|}
\hline
\# & $r_{E},\sigma_{E}\times10^{4}$ & $r_{G},\sigma_{G}\times10^{4}$ & $%
r_{PL},\sigma_{PL}\times10^{4}$ & $\sigma_{n}\times10^{17}$ & $%
\Phi_{E},\Phi_{G},\Phi_{PL},\Phi_{n}$ \\ \hline
1 & 1.27, 6.69 & 0.85, 6.93 & 1.26, 6.72 & 3.79 & 18.6, 27.4, 19.6, 25.8 \\ 
\hline
2 & 1.23, 6.83 & 0.88, 6.94 & 1.25, 6.84 & 1.49 & 19.1, 26.8, 19.7, 26.2 \\ 
\hline
3 & 1.04, 6.51 & 0.73, 6.74 & 1.07, 6.54 & 2.82 & 20.7, 30.2, 21.4,27.3 \\ 
\hline
4 & 1.18, 6.65 & 0.87, 6.71 & 1.23, 6.62 & 3.01 & 19.7, 27.1, 20.0, 26.1 \\ 
\hline
5 & 0.94, 6.44 & 0.74, 6.42 & 1.03, 6.38 & 1.91 & 22.2, 29.8, 21.9, 27.7 \\ 
\hline
\end{tabular}
\end{center}

\textbf{Table III. }Five Monte Carlo runs for the $\mathrm{1D}$ Ising model.
The "true" correlation function is exponential with $r=1.19$ and $\eta=0.119$
and yields $\Phi_{E}^{th}=19.5$. The correlation functions extracted from
the generated rough surfaces were fitted with the exponential, Gaussian, and
power law functions. The Table contains the best fitting values of $%
r_{E,G,PL}$ and the corresponding values of $\sigma_{E,G,PL}$\ and $%
\Phi_{E,G,PL}$. Since the simulation is based on the Ising model with spins $%
\pm1$, the best fitting values of $\eta$ differed from $0.119$ by less than
1\% for all fitting functions. The values of $\Phi_{n}$ were obtained by
direct spectral analysis of the raw correlation data. The size of the sample
was $N=1000$ and we performed $10^{6}$ Metropolis cycles.

The last table, Table IV, presents results for three rough surfaces
generated using the $\mathrm{2D}$ Ising model plus a row for the correlation
function averaged over ten runs. The observable here is again $\zeta_{0}$.

\begin{center}
\begin{tabular}{|c|c|c|c|c|c|}
\hline
\# & $r_{E1},\sigma_{E1}\times10^{2}$ & $r_{E2},\sigma_{E2}\times10^{2}$ & $%
r_{G},\sigma_{G}\times10^{2}$ & $r_{PL},\sigma_{PL}\times10^{2}$ & $\zeta
_{0}^{E1},\zeta_{0}^{E2},\zeta_{0}^{G},\zeta_{0}^{PL}$ \\ \hline
1 & 1.56, 2.03 & 1.60, 2.75 & 1.06, 2.41 & 1.55, 2.12 & 15.33, 16.18, 7.01,
15.16 \\ \hline
2 & 1.43, 1.56 & 1.43, 2.27 & 1.06, 1.89 & 1.48, 1.63 & 12.80, 12.94, 7.11,
13.78 \\ \hline
3 & 1.53, 1.66 & 1.53, 2.49 & 1.11, 2.04 & 1.57, 1.75 & 14.61, 14.75, 7.80,
15.43 \\ \hline
4 & 1.54, 0.69 & 1.57, 0.89 & 1.10, 1.42 & 1.57, 0.91 & 14.99, 15.43, 7.67,
15.46 \\ \hline
\end{tabular}
\end{center}

\textbf{Table IV. }Results for three rough surfaces generated using the $%
\mathrm{2D}$ Ising model (the first three rows) and for the correlation
function averaged over ten runs (the last row). The Monte Carlo simulations
have been done at $T=1.2T_{c}$ with $10^{6}$ Metropolis cycles as in Figure
7. The surface size is $100\times100$. The Table is arranged similarly to
Table II. The Table contains the best fitting values of $r_{E,G,PL}$ and the
corresponding values of $\sigma_{E,G,PL}$\ and $\zeta_{0}^{E,G,PL}$. The
values of $\zeta_{0}^{n1,2}$ for direct spectral analysis of the raw
correlation data are given in the text. The results for the exponential fits 
$\zeta_{0}^{E1,2}$ for $\zeta\left( \left\vert \mathbf{s}\right\vert \right) 
$ and $\zeta\left( \mathbf{s}\right) $ should be the closest to the true
physical parameters.

The computations are done above the phase transition, $T=1.2T_{c}$. At this
temperature the correlation function is, probably, still close to the
exponential (Figures 7$b$ and 7$c$), but it is not clear how close. Here we
do not know exactly what should be the "true" value of $\zeta _{0}$, but
expect that the exponential correlator provides the best estimate. At this
temperature the domains are relatively small (see Figure 7$a$) and the
relaxation times are manageable. The size of the surface is relatively
large, $100\times 100$, and each computation runs $10^{6}$ Metropolis
cycles. The Table is arranged similarly to Table II. Here again the values
of $\sigma $ for all fitting functions are close for each other while the
values of $\zeta _{0}$ and $r$ are noticeably different. The results for the
exponential fit should be the closest to the true physical parameters. The
first column for the exponential fitting gives results obtained from the
flat ($\mathrm{1D}$) file $\zeta \left( \left\vert \mathbf{s}\right\vert
\right) $. The second column gives the results of fitting $\zeta \left( 
\mathbf{s}\right) $ by the $\mathrm{2D}$ exponential function. For the
Gaussian and power law correlators, columns 3 and 4, we used only the flat
files $\zeta \left( \left\vert \mathbf{s}\right\vert \right) $.What is
somewhat surprising is that the results for our choice of the power law
correlator, which is the Fourier image of the exponential one, are again
close to those using the exponential fit. What is even more surprising, the
values of $\zeta _{0}$ for the power law fit using \textrm{1D} $\zeta \left(
\left\vert \mathbf{s}\right\vert \right) $\ are systematically closer to the
exponential fit using \textrm{2D} $\zeta \left( \mathbf{s}\right) $ than to
the exponential fit using \textrm{1D} $\zeta \left( \left\vert \mathbf{s}%
\right\vert \right) $. The Gaussian fit yields very different $\zeta _{0}$
while the value of $\sigma $ is comparable with the others. The direct
spectral analysis of the raw correlator data again yields the worst physical
results and changes from run to run; there results are not even worth
listing. The spectral analysis of the correlation function averaged over ten
runs worked better than the Gaussian fit and yielded $\zeta _{0}^{n1}=17.42$
for the flat files $\zeta \left( \left\vert \mathbf{s}\right\vert \right) $
and $\zeta _{0}^{n1}=18.70$ when working with the \textrm{2D} correlation
function $\zeta \left( \mathbf{s}\right) $. The differences between results
obtained using different fitting functions once again illustrate the
uncertainty in comparing computational and experimental data to theoretical
results. One should have at least some information about the shape of the
"true" correlation function.

\section{Summary and conclusions}

In summary, we looked at reconciling numerical and physical measurements of
random rough surfaces with theoretical results using the roughness
correlation function. We demonstrated that data extracted from scanning
microscopy of the surface profile can be insufficient for unambiguous
determination of the shape of the correlation function (for some of the
recent experimental attempts to extract the correlation function from the
scanning microscopy see Refs. \cite%
{munoz1,munoz4,shikin1,munoz2,munoz3,parpia1}). The same is true for
computational experiments in which a random surface is generated without an
effort to reproduce a known correlation function.

There are two main obstacles apart from the accuracy of measurements. The
first one is the presence of fluctuations which are unavoidable for finite
samples. The second one is the relationship between the step size $b$,
correlation radius $r$, and the overall number of data points $N$. To
properly recover the shape of the correlation maximum, one needs the step
size $b$\ to be noticeably smaller than the correlation radius $r$. If one
decreases $b$ (or, what is the same, increases $r$) while keeping the
overall number of data points $N$, which is determined by the technical or
computational abilities, constant, the data set measured in units of $r$
effectively shrinks. Since $r$ determines the size of correlated clusters
(domains), the full data set will cover the smaller number of domains. This,
in turn, gives rise to noticeable spurious, purely geometrical interdomain
correlations which have nothing to do with real physical interactions. These
interdomain correlations can masquerade as the presence of an additional,
larger correlation radius. The same effect makes reproducing surfaces with
very large correlation radii virtually impossible.

We analyzed two methods for numerical generation of surfaces with
predetermined roughness correlation functions. This was done with practical
physical applications in mind: \textrm{1D} beams of ultracold neutrons in a
rough waveguide, resistivity of ultrathin rough films in quantum size effect
conditions, and particle or wave scattering in the longwave limit. We judged
the quality of the analysis of the extracted correlation functions by the
accuracy of the predictions for observables for these applications.

For the neutron problem, for which the roughness of the waveguide is
introduced on purpose, we suggest a practical way of preparing the rough
mirror for optimization of the GRANIT-type experiment \cite%
{neutron1,nesv2,mauricio1}. Our recommendation is to generate a random rough
profile using the Monte Carlo simulations on the basis of the $\mathrm{1D}$
Ising model with the correlation radius $r=2$ and the amplitude of roughness
in the $0.2\div0.4$ range and to transfer this profile onto the mirror
surface. This allows one to increase the value of $\Phi$ to $43.5\div170$, 
\emph{i.e.}, several times times in comparison to what it is assumed to be
in previous experiments while creating a perfectly controllable environment.
This is sufficient for showing the well-developed quantum steps in the exit
neutron count and produce neutrons with well-defined energies in the \textrm{%
peV} range. Since all the lengths here are in the units of \textrm{6 }$%
\mathrm{\mu m}$, this procedure seems to be feasible.

There are several challenges for identifying the roughness correlator from
experimental and numerical data on the surface profile even if one
disregards all the issues concerning the accuracy of profile measurements.
Most importantly, the standard deviation $\sigma$ between the measured or
generated correlation function and some fitting function cannot be
considered a good predictor for physical results.

The value of $\sigma$ extracted from fitting is strongly weighted towards
the tail of the correlation function. If the correlator rapidly decreases at
large distances, the values of $\sigma$ are more or less the same for all
reasonable fitting functions and measure the fluctuations without saying
anything about appropriateness of the chosen fitting functions. Meanwhile,
the physical observables are very sensitive to the shape of the correlator.
As a result, the error in physical results can by far exceed $\sigma$.

Decreasing the size of the fluctuations requires increasing the size of a
sample. Increasing the size of the sample, on the other hand, increases the
role of the fluctuation-driven tails of the correlators at the expense of
the contribution from the peak area in which one would expect to observe
main differences between the physically different correlators.

One option for suppressing the fluctuation-driven tails is to average the
numerically or experimentally measured correlation function over several
samples as it is sometimes done in experiment \cite{munoz2}. However, this
operation can be inherently\ dangerous when, for example, the correlation
function itself has longer tails of alternating sign. If one knows that
there are no long range correlations, this averaging over several samples
can be very helpful for \textrm{2D} roughness. Such averaging has not been
necessary for \textrm{1D} roughness in our numerical experiments. The same
difficulty persists if one simply cuts off the long range tails assuming
that they are driven only by the fluctuations.

Generating or measuring the correlation function with a large correlation
radius $R$ requires a noticeable increase in the sample size $N$: the
important parameter is not the overall number of the data points $N$, but
the number of inhomogeneities $N/N_{i}$ where $N_{i}$ is the number of
points in a typical inhomogeneity. The problem is exacerbated in the
two-dimensional case when $N_{i}$ grows proportionally to $R^{2}$. The shape
of the correlation function with not very large $N/N_{i}$ could be very
misleading and point, rather convincingly, at fictitious long-range
correlations.

In general, it is much easier to generate a rough surface with a desired
correlator in a $\mathrm{1D}$ rather than in a $\mathrm{2D}$ case. Apart
from the obvious difficulty that the $\mathrm{2D}$ sample of the same linear
size $L$ requires the use of $N=L^{2}$ data points rather than $N=L$ as in
the $\mathrm{1D}$ case, there is an additional difficulty associated with a
greater volatility of the correlation function due to the residual
anisotropy in generated or measured correlators.

We also tried the alternative approach to data analysis without fitting
functions by performing the spectral analysis of the raw correlation data
and using the results for direct calculation of observables. In \textrm{1D}
examples this approach worked somewhat, but not much, better than using a
fitting function of a wrong shape, but still noticeably worse than using the
"right" fitting function. This approach did not work for us in \textrm{2D }%
cases because of the fluctuation-driven anisotropy of the extracted
correlators and smaller linear sizes of the samples than in \textrm{1D}.

If there are no restriction on the amplitudes of inhomogeneities, as in the
case of macroscopic roughness, one can easily generate a surface with any
given correlation function. Generating random surfaces with discretized
(atomic) inhomogeneities, \textit{i.e.}, inhomogeneities with amplitudes of
integer sizes, presents unique challenges. Here the only reliable method is
to use a solvable lattice model (for example, the Ising model). The universe
of exactly solvable models is limited and, therefore, one can generate the
surfaces with discrete amplitudes of inhomogeneities with just few types of
the predetermined surface correlators which may or may not reflect the real
rough surfaces.

\begin{acknowledgement}
\textit{We are grateful to P. Nightingale for helpful discussions concerning
generation of random surfaces. }One of the authors (A.M.) is grateful to the
members of GRANIT group at ILL, and especially to V. Nesvizhevsky, for the
hospitality during his visits to Grenoble and for the stimulating
discussions.
\end{acknowledgement}

\section{Figure captions}

\textbf{Figure 1.} (color online) An example of the correlation function
(black solid line) for a generated $\mathrm{1D}$ surface which should
emulate a surface with Gaussian correlation of inhomogeneities $\zeta \left(
x\right) =\exp \left( -x^{2}/8\right) $ (blue dashed line). The total number
of points is 2000, the average amplitude of roughness $\eta =\ell /l_{0}=1$,
the correlation radius $r=R/l_{0}=2.$

\textbf{Figure 2. }(color online)\textbf{\ }Correlation functions for $%
\mathrm{1D}$ generated surface profiles which should emulate the Gaussian
(black; curve 1), exponential (red; curve 2), and PL (blue; curve 3)
correlation functions. In the peak area (Figure $2a$) the differences are
very pronounced, but the fluctuation-driven tails (Figure $2b$) are almost
identical. All three computations started from the same set of $N=2000$
random numbers.

\textbf{Figure 3}. (color online) Dependence of the standard deviation $%
\sigma$ between generated and exact correlation functions on the sample size 
$N$. The solid line is $\sqrt{2/N}$. The generated roughness is supposed to
have Gaussian correlations with $r=2$.

\textbf{Figure 4.} (color online) An example of a $\mathrm{2D}$ rough
surface of the size $60\times 60$. The roughness emulates isotropic Gaussian
correlations with $r=2$, $\zeta \left( s\right) =\exp \left( -s^{2}/8\right) 
$ (black line 1 in Figure 4$b$). ($a$) $2D$ correlation function $\zeta
\left( x,y\right) $ ($b$) The correlation function $\zeta \left( s\right) $
after averaging over the angles (line 2; blue).

\textbf{Figure 5.} (color online) Correlation function for a generated
surface emulating $\zeta \left( i-k\right) =4\exp \left( -\left\vert
i-k\right\vert /2\right) $ (line 1; black) after rounding the profile data
points $y_{i}$ to the nearest integer number $\overline{y}_{i}$. Line 2
(red): the generated raw correlator $\zeta \left( \left\vert i-k\right\vert
\right) =\left\langle y_{i}y_{k}\right\rangle $, line 3 (blue): the
correlator for the discretized surface $\left\langle \overline{y}_{i}%
\overline{y}_{k}\right\rangle $.

\textbf{Figure 6.} (color online) An example of the correlation function for
a $\mathrm{1D}$ rough surface $y\left( x\right) $ and its correlation
function $\zeta \left( s\right) $ generated using the Ising model (red line
2). The parameters in the correlation function $r=1.19$ and $\eta =0.119$;
black line 1 is given by Eq. $\left( \ref{a1}\right) $ ($\sigma =6.83\times
10^{-4}$).

\textbf{Figure 7.} (color online) An example of a $\mathrm{2D}$\ random
rough surface generated using the Ising model at $T=1.2T_{c}$: $a)$ \textrm{%
2D }correlation function $b)$ correlation function after averaging over the
angles.

\section{Table captions}

\textbf{Table I.} Three numerical runs based on Sec. IIA in application to
our neutron problem. Rough $1D$\ surfaces emulate Gaussian correlation of
inhomogeneities $\eta ^{2}\exp \left( -x^{2}/2r^{2}\right) $ with $r=1.19$
and $\eta =0.119$ as it was assumed in experiment. The expected value of the
main physical parameter $\Phi $, Eq. $\left( \ref{ne2}\right) $, is $\Phi
=23.48$. The extracted correlators were fitted with Gaussian, $\eta
_{G}^{fit}\exp \left( -s^{2}/2r_{G}^{fit}\right) $, exponential, $\eta
_{E}^{fit}\exp \left( -s/r_{E}^{fit}\right) $, and power law, $\eta
_{PL}^{fit}/\left[ 1+\left( s/r_{PL}^{fit}\right) ^{2}\right] ^{3/2}$
fitting functions. The table contains the best fitting values of $%
r_{G,E,PL}^{fit}$, together with $\sigma _{G,E,PL}$, and the recalculated
values of $\Phi _{G,E,PL}$. The columns with $\Phi _{n}$ and $\sigma _{n}$
give the values of $\Phi $ and the standard deviation when the spectral
decomposition of the data was put directly into equations without fitting.

\textbf{Table II.} The same as in Table I for generated $2D$ rough Gaussian
surfaces with $r=2$ and $\eta=1$. The expected value of the main physical
observable $\zeta\left( q=0\right) =8\pi.$ The table contains the extracted
fitting parameters $\eta_{G,E,PL}^{fit}$ and $r_{G,E,PL}^{fit}$, together
with $\sigma_{G,E,PL}$, and the recalculated values of $\zeta_{0}$. The
Gaussian fit was done independently for the \textrm{1D} correlation function 
$\zeta\left( \left\vert \mathbf{s}\right\vert \right) $ (index 1) and the 
\textrm{2D} correlation function $\zeta\left( \mathbf{s}\right) $ (index 2).
The fourth row gives the results for the correlation function averaged over
10 independent runs.

\textbf{Table III. }Five Monte Carlo runs for the $\mathrm{1D}$ Ising model.
The "true" correlation function is exponential with $r=1.19$ and $\eta=0.119$
and yields $\Phi_{E}^{th}=19.5$. The correlation functions extracted from
the generated rough surfaces were fitted with the exponential, Gaussian, and
power law functions. The Table contains the best fitting values of $%
r_{E,G,PL}$ and the corresponding values of $\sigma_{E,G,PL}$\ and $%
\Phi_{E,G,PL}$. Since the simulation is based on the Ising model with spins $%
\pm1$, the best fitting values of $\eta$ differed from $0.119$ by less than
1\% for all fitting functions. The values of $\Phi_{n}$ were obtained by
direct spectral analysis of the raw correlation data. The size of the sample
was $N=1000$ and we performed $10^{6}$ Metropolis cycles.

\textbf{Table IV. }Results for three rough surfaces generated using the $%
\mathrm{2D}$ Ising model (the first three rows) and for the correlation
function averaged over ten runs (the last row). The Monte Carlo simulations
have been done at $T=1.2T_{c}$ with $10^{6}$ Metropolis cycles as in Figure
7. The surface size is $100\times100$. The Table is arranged similarly to
Table II. The Table contains the best fitting values of $r_{E,G,PL}$ and the
corresponding values of $\sigma_{E,G,PL}$\ and $\zeta_{0}^{E,G,PL}$. The
values of $\zeta_{0}^{n1,2}$ for direct spectral analysis of the raw
correlation data are given in the text. The results for the exponential fits 
$\zeta_{0}^{E1,2}$ for $\zeta\left( \left\vert \mathbf{s}\right\vert \right) 
$ and $\zeta\left( \mathbf{s}\right) $ should be the closest to the true
physical parameters.


\newpage

\begin{thebibliography}{99}
\bibitem{q2} J. A. Ogilvy, \textit{Theory of Wave Scattering from Random
Surfaces} (Adam Hilger, Bristol) 1991

\bibitem{bernasek1} P. Dooley, and S.L. Bernasek, Surf. Sc. \textbf{406} 206
(1998)

\bibitem{fracture1} B. B. Mandelbrot, D. E. Passoja, and A. J. Paullay,
Nature (London) \textbf{308}, 721 (1984)

\bibitem{fracture2} L. Ponson, D. Bonamy, and E. Bouchaud, Phys. Rev. Lett. 
\textbf{96}, 035506 (2006)

\bibitem{fracture3} Jan Oystein Haavig Bakke, and A. Hansen, Phys. Rev. E 
\textbf{76}, 031136 (2007)

\bibitem{fer1} R. M. Feenstra, D. A. Collins, D. Z.-Y. Ting, M. W. Wang, and
T. C. McGill, Phys. Rev. Lett. \textbf{72}, 2749 (1994)

\bibitem{munoz1} R. C. Munoz, G. Vidal, G. Kremer, L. Moraga, C. Arenas, and
A. Concha, J. Phys.: Condens. Matt., \textbf{12}, 2903 (2000)

\bibitem{pon1} A. E. Meyerovich, and I. V. Ponomarev, Phys.Rev. B \textbf{65}%
, 155413 (2002); Y. Cheng, and A. E. Meyerovich, Phys.Rev. B \textbf{73,}
085404 (2006)

\bibitem{stm1} \textit{Scanning Tunneling Microscopy}, Eds. J. A. Stroscio,
and W. J. Kaiser (Academic Press, NY) 1993 [Methods of Experimental Physics,
v. 27]

\bibitem{munoz4} R. C. Munoz, G. Vidal, M. Mulsow, J. G. Lisoni, C. Arenas,
A. Concha, F. Mora, R. Espejo, G. Kremer, L. Moraga, R. Esparza, and P.
Haberle, Phys. Rev. B \textbf{62}, 4686 (2000)

\bibitem{ogilvy2} J A Ogilvy, and J R Foster, J. Phys. D: Appl. Phys, 
\textbf{22}, 1243 (1989)

\bibitem{shikin1} A. Fubel, M. Zech, P. Leiderer, J. Klier, and V. Shikin,
Surf. Sc. \textbf{601}, 1684 (2007)

\bibitem{goodnick1} S. M. Goodnick, D. K. Ferry, C. W. Wilmsen, Z.
Liliental, D. Fathy, and O. L. Krivanek, Phys. Rev. B \textbf{32}, 8171
(1985)

\bibitem{palas1} G. Palasantzas, Phys. Rev. B \textbf{48}, 14472 (1993)

\bibitem{munoz2} M. E. Robles, C. A. Gonzalez-Fuentes, R. Henriquez, G.
Kremerc, L. Moragab, S. Oyarzunb, M. A. Suarez, M. Flores, R. C. Munoz,
Appl. Surf. Sc. \textbf{258}, 3393--3404 (2012)

\bibitem{tribology1} Jiunn-Jong Wu, Tribology Internat. \textbf{33} 47 (2000)

\bibitem{navarini1} J. C. Novarini, and J. W. Caruthers, J. Acoust. Soc. Am. 
\textbf{53}, 876 (1973)

\bibitem{johnson1} J. D. Johnson, S. Krinsky, and B. M. McCoy, Phys. Rev. 
\textbf{A8}, 2526 (1973); Hung Cheng and Tai Tsun Wu, Phys. Rev. \textbf{164}%
, 719 (1967)

\bibitem{neutron1} V. V. Nesvizhevsky \textit{et al,} Nature, \textbf{415},
297 (2002); V. V. Nesvizhevsky \textit{et al,} Phys. Rev. D \textbf{67},
102002 1-9 (2003); V.V.Nesvizhevsky \textit{et al}, European Phys. Journal C 
\textbf{40,} 479 (2005) [for additional bibliography see also
http://lpscwww.in2p3.fr/UCN/NiveauxQ\_G/publications/index.html]

\bibitem{nesv2} V. V. Nesvizhevsky, Physics - Uspekhi, \textbf{53}, 645
(2010)

\bibitem{breit1} G. Breit, Phys. Rev., \textbf{32}, 273 (1928)

\bibitem{freed1} J. H. Freed, Ann. Phys. Fr., \textbf{10}, 901 (1985)

\bibitem{pdg1} H. Murayama, G. G. Raffelt, C. Hagmann, K. van Bibber, and
L.J.Rosenberg, in: K. Hagiwara et al. (Particle Data Group), \textit{Review
of Particle Properties}, p. 374, Phys. Rev. D \textbf{66}, 010001 (2002)

\bibitem{prot1} V. V. Nesvizhevsky, and K. V. Protasov, Class. Quantum
Gravity, \textbf{21}, 4557 (2004)

\bibitem{grenoble1} Possible use of gravitational quantization of neutrons
for measuring the fundamental forces was the main topic of the workshop
http://lpsc.in2p3.fr/congres/granit06/index.php (ILL, Grenoble, April 2006)

\bibitem{mirror} V.V. Nesvizhevsky \textit{et al,} Nuclear Instruments and
Methods in Physics Research (NIM) A \textbf{578}, 435 (2007)

\bibitem{mauricio1} M. Escobar, and A. E. Meyerovich, Phys. Rev. A \textbf{83%
}, 033618 (2011)

\bibitem{arm2} A. E. Meyerovich, and A. Stepaniants, Phys. Rev. B \textbf{60}%
, 9129 (1999)

\bibitem{munoz3} R. C. Munoz, G. Vidal, G. Kremer, L. Moraga, C. Arenas, J.
Phys.: Condens. Matt. \textbf{11}, L299 (1999)

\bibitem{parpia1} P. Sharma, A. C\'{o}rcoles, R. G. Bennett, J. M. Parpia,
B. Cowan, A. Casey, and J. Saunders, Phys. Rev. Lett. \textbf{107}, 196805
(2011)\newpage
\end{thebibliography}
\end{document}